\documentclass[aps,prx,twocolumn,superscriptaddress,nofootinbib]{revtex4-2}

\usepackage{booktabs}
\usepackage{feyn}
\usepackage{graphicx}
\usepackage{latexsym}
\usepackage{amssymb}
\usepackage{amsmath}
\usepackage{amsthm}
\usepackage{mathdots}
\usepackage{amsfonts}
\usepackage{bm}
\usepackage{multirow}
\usepackage{color}
\usepackage{comment}
\usepackage{bbm}
\usepackage{marvosym}
\usepackage{wasysym}
\usepackage{enumerate}  
\usepackage{hyperref}
\usepackage{url}
\usepackage{orcidlink}
\usepackage{mathrsfs}
\usepackage{CJK}
\usepackage{titletoc}
\usepackage{tikz}
\usepackage{quantikz}
\usepackage[table]{xcolor}
\newcommand{\ii}{\mathrm{i}}

\newcommand{\id}{\mathbbm{1}}
\newcommand{\Dcal}{\mathcal{D}}

\newcommand{\Uenc}{U_{\rm enc}}

\definecolor{cinnabar}{rgb}{0.89, 0.26, 0.2}

\newcounter{para}

\usepackage{array, makecell}

\newcommand{\xCornell}{Department of Physics, Cornell University, Ithaca, NY, USA}

\begin{document}
\begin{CJK*}{UTF8}{}

\title{Cultivating logical catalysts for fault-tolerant dyadic phase rotations}

\author{Yichen Xu \CJKfamily{gbsn}(许轶臣)~\orcidlink{0000-0001-7914-9438}}
\thanks{yx639@cornell.edu}
\affiliation{\xCornell}

\author{Xiao Wang \CJKfamily{gbsn}(王骁)~\orcidlink{0000-0003-2898-3355}}
\thanks{xw753@cornell.edu}
\affiliation{\xCornell}

\begin{abstract}
We introduce a surface-code cultivation protocol for reusable logical catalyst states that implement exact fine dyadic phase gates \(Z^{2^{-b}}\) by phase kickback.  
The catalyst is an eigenstate of a high-period Clifford circuit \(U\), with a direct construction supported on \(O(2^b)\) logical qubits.  
Once cultivated, each invocation implements the target phase through a controlled-\(U\) gadget, removing Clifford+\(T\) synthesis approximation error from the online gate and making the online non-Clifford depth independent of the target logical accuracy.  
As a concrete demonstration, we construct a catalyst for \(\sqrt{T}=Z^{1/8}\), where \(U\) is a nine-qubit brickwork Clifford circuit and controlled-\(U\) consists of eight controlled-CNOTs. 
Starting from nine distance-three rotated-surface-code blocks, we cultivate the catalyst through logical-\(U\) checks, syndrome extraction and postselection, code growth, and complementary-gap decoding.  
Due to the intrinsic fault tolerance of the phase read-out, a \emph{single} verification round already reaches the leading error-corrected scaling, in contrast to the repeated logical checks required when cultivating single-qubit magic states.  
A hybrid tensor-network and stabilizer simulation shows that, at physical error rate \(p=10^{-3}\), the postselected catalyst can be grown to distance-seven rotated-surface-code blocks with logical leakage rate \(\sim 10^{-6}\) using around seven expected attempts, and can be suppressed further with stronger postselection.  
Compared with existing protocols, our approach trades offline, phase-specific catalyst cultivation for exactness, reusability, and constant-depth online implementation of fixed fine dyadic phases in codes with restricted transversal gate sets.
\end{abstract}
 
\maketitle
\end{CJK*}
\tableofcontents
 
\section{Introduction}
 
Fault-tolerant quantum computation~\cite{shor1996fault,aharonov2008fault,knill1998resilient} requires resources beyond stabilizer operations, since Clifford circuits acting on stabilizer states are efficiently simulable classically~\cite{gottesman1998heisenberg,aaronson2004improved}.
In codes whose native transversal gates are restricted, such as the rotated surface code~\cite{fowler2012surface}, the standard route to universality is to supplement Clifford operations with non-Clifford resource states.
The most widely used example is the $T=Z^{1/4}$ gate, supplied by injected or distilled magic states~\cite{bravyi2005universal}\footnote{Throughout this work we use the convention $Z^\alpha = \operatorname{diag}(1,e^{\ii\pi\alpha})$, so that $\sqrt{T}=Z^{1/8}$.}.
Within Clifford+$T$, rotations that are not exactly synthesizable must be approximated: the Solovay--Kitaev theorem gives a general polylogarithmic approximation algorithm~\cite{dawson2006solovay}, while number-theoretic synthesis achieves near-optimal $T$-counts for $Z$-rotations, scaling as $3\log_2(1/\varepsilon)+O(\log\log(1/\varepsilon))$ in the ancilla-free setting~\cite{kliuchnikov2013fast,ross2016optimal}.
Repeat-until-success circuits reduce the constant factor by using measurements, ancillas, and retries~\cite{paetznick2014repeat,bocharov2015efficient}.
In all of these approaches, however, the synthesis precision $\varepsilon$ must be chosen below the target logical error rate of the code, so the non-Clifford cost grows as the computation is pushed to lower logical error.
 
A complementary strategy is to prepare specialized non-Clifford resource states directly, rather than approximating every target gate from a fixed elementary library.
This perspective includes resource states for Toffoli, $\mathrm{CCZ}$, and small-angle rotations~\cite{jones2013low,eastin2013distilling,campbell2017unifying,campbell2016efficient,duclos2015reducing}, and it has led to several fault-tolerant preparation paradigms.
Magic state distillation~\cite{bravyi2005universal} consumes many noisy copies of a target magic state and outputs fewer, higher-fidelity copies~\cite{bravyi2012magic,haah2018codes,litinski2019magic,litinski2019game,gidney2019efficient}; logical distillation has now also been demonstrated experimentally~\cite{rodriguez2025experimental}.
Code switching and gauge fixing move encoded states between codes with complementary transversal gate sets~\cite{bombin2015gauge,anderson2014fault,paetznick2013universal,kubica2015universal,bombin2016dimensional,beverland2021cost,butt2024fault,heusen2025efficient}.
More recently, magic state cultivation prepares a single low-distance logical magic state, verifies it by logical checks, and grows it to a larger code distance while postselecting on successful outcomes~\cite{gidney2024magic,itogawa2024even,hirano2024leveraging}.
This idea has quickly developed into several surface-code and related architectures~\cite{vaknin2025efficient,chen2025rp2,claes2025cultivating,sahay2025fold,hirano2025lattice,rosenfeld2025magic,chen2026efficient}.
Existing cultivation protocols, however, primarily target the canonical $\ket{T}\equiv T|+\rangle$ state, which is consumed once per non-Clifford operation.
 
Directly cultivating finer dyadic magic states is obstructed by the Clifford hierarchy~\cite{gottesman1999demonstrating}.
The state $|T\rangle $ is stabilized by $TXT^\dagger$, which is a Clifford operator.
As a result, its low-distance verification can be based on logical Clifford checks.
For a finer dyadic magic state $Z^{2^{-b}}|+\rangle$ with $b\geq 3$, the analogous stabilizer $Z^{2^{-b}}XZ^{-2^{-b}}$ lies at a higher level of the hierarchy and is non-Clifford.
A direct verification of such a state would therefore require precisely the kind of non-Clifford measurement resource that the protocol is meant to produce.
This is the basic reason that the successful cultivation paradigm for $|T\rangle$ does not immediately extend to $\sqrt{T}$ or to still finer dyadic phases. One could, in principle, start from a small instance of a code with transversal $Z^{2^{-b}}XZ^{-2^{-b}}$ gates. However, the subsequent growth to a surface-code or color-code architecture requires intricate grafting, as demonstrated by a recent work~\cite{chen2026efficient}.
 
Catalyst states address a different bottleneck: the per-use cost of rotations.
A catalyst is returned to itself after the gate implementation, so its preparation cost can be amortized over many invocations.
Phase kickback from Kitaev's phase-estimation framework~\cite{kitaev1995quantum,kitaev2002classical} is the canonical mechanism.
The phase-gradient state~\cite{kitaev1995quantum,gidney2018halving}, for example, converts controlled additions into dyadic rotations and is a central primitive in modern fault-tolerant resource estimates for Shor's algorithm~\cite{gidney2021factor,gidney2025factor} and Hamiltonian simulation~\cite{babbush2018encoding,lee2021even}.
Phase-gradient methods are qubit-efficient, using $O(b)$ logical qubits for $b$ bits of phase precision, but their online implementation uses controlled arithmetic whose non-Clifford cost and depth scale with $b$ in standard constructions.
A recent algebraic approach by Kim gives catalytic $z$-rotations in constant $T$-depth for arbitrary target angles, assuming access to suitable catalyst eigenstates of finite-field Clifford operators~\cite{kim2025catalytic}.
That construction is powerful asymptotically: the catalyst size is polynomial in $\log(1/\varepsilon)$ for an $\varepsilon$-approximate rotation, but its accessible rotation angles are of the form $2\pi k/(2^n - 1)$, which are not dyadic. Representing $Z^{2^{-b}}$ exactly is therefore left as an open problem, and the fault-tolerant preparation of the catalyst itself involves nontrivial finite-field and phase-estimation structure rather than a low-distance-to-high-distance cultivation protocol.
 
In this work, we combine the amortization advantage of catalysts with the code-growth philosophy of cultivation, and resolve the dyadic-exactness gap for a representative fixed angle.
Our starting point is the controlled-jump rule for the Clifford hierarchy~\cite{xu2026controlled}: if a Clifford unitary $U$ has Pauli period $m$, meaning that $U^{2^m}$ is a Pauli operator up to phase, then the controlled-$U$ gate lies at level $m+2$ of the hierarchy.
An eigenstate of such a high-period Clifford circuit can therefore be used as a reusable catalyst: controlled-$U$ phase kickback implements a fine dyadic phase \emph{exactly} at the algebraic level, while the online non-Clifford cost is governed by the structured controlled-Clifford circuit rather than by a Clifford+$T$ approximation length.
The tradeoff is distinct from both phase-gradient catalysts and Kim's algebraic catalysts.
Our direct construction uses a phase-specific catalyst supported on $O(2^b)$ logical qubits, but for fixed $b$ it gives an exact dyadic phase with constant online non-Clifford depth and a concrete surface-code cultivation pathway.
The online invocation cost is therefore exact at the algebraic level: no rotation-synthesis approximation error is introduced when the gate is used, and bounded only by the residual logical infidelity of the cultivated catalyst.
 
We demonstrate this idea for a $\sqrt{T}=Z^{1/8}$ catalyst, an eigenstate of a nine-qubit brickwork CNOT circuit $U_9$ with eigenvalue $e^{\ii\pi/8}$.
The cultivation protocol prepares a physical nine-qubit eigenstate, encodes its qubits into nine independent distance-three rotated surface-code blocks, and verifies the encoded catalyst with a logical-$U_9$ phase-estimation measurement; runs that pass with trivial distance-three syndromes are grown to distance-seven blocks through unitary growth, stabilizer-measurement growth, and complementary-gap decoding.
A key feature of this construction is that the logical-$U_9$ check is intrinsically fault-tolerant, since its exact phase read-out catches any single fault that corrupts the catalyst's eigenphase. As a result, a \emph{single} verification round already suffices, without the repeated logical measurements that $\ket{T}$-cultivation protocols require.
 
Because the protocol consumes a substantial number of non-Clifford gates during verification, we evaluate it with a hybrid simulation.
A tensor-network state-vector simulation tracks the logical state through the noisy logical-$U_9$ verification, and the ancilla cleanup measurements; 
the postselected handoff is then passed to a Stim-based stabilizer simulation~\cite{gidney_stim_2021} with matching-based decoding via PyMatching~\cite{higgott2022pymatching}.
At physical error rate $p=10^{-3}$, the catalyst is grown to distance-seven blocks with a logical leakage rate $\sim10^{-6}$ in roughly seven expected attempts, falling toward $\sim10^{-7}$ under stronger complementary-gap postselection.
This matches the logical infidelity of recent single-block $\ket{T}$-cultivation protocols~\cite{sahay2025fold,claes2025cultivating}, despite our catalyst supporting a finer dyadic phase on nine surface-code blocks rather than one.
 
The rest of the paper is organized as follows.
In Sec.~\ref{sec:protocol} we present the controlled-Clifford catalyst construction and the surface-code cultivation circuit.
In Sec.~\ref{sec:simulation} we describe the hybrid tensor-network and stabilizer simulation and report the cultivation performance.
Finally, in Sec.~\ref{sec:application}, we compare the resource tradeoffs of our protocol with Clifford+$T$ synthesis, repeat-until-success circuits, and phase-gradient catalysts.
 
\section{Cultivation circuit of a logical catalyst state}\label{sec:protocol}
 
This section lays out the theoretical foundation of our protocol in three steps.
Section~\ref{sec:protocol-catalyst} introduces the controlled-Clifford catalyst mechanism for general high-period brickwork CNOT circuits and shows how eigenstate phase kickback realizes a dyadic rotation exactly.
Section~\ref{sec:protocol-sqrtT} specializes to nine qubits, where the resulting catalyst implements $\sqrt{T}$ with a small constant magic-state cost per invocation.
Section~\ref{sec:protocol-cultivation} then presents the surface-code cultivation circuit that fault-tolerantly prepares this catalyst, organized around three stages: logical-$U_9$ verification by phase estimation, syndrome-based postselection at distance three, and code growth to distance seven with complementary-gap decoding.
 
\subsection{Controlled-Clifford catalysts from high-period CNOT circuits}
\label{sec:protocol-catalyst}
 
We first review the controlled-Clifford catalyst mechanism for fine dyadic phase gates, first proposed in Ref.~\cite{xu2026controlled}. 
We use the convention $\mathrm{CNOT}_{i\to j}$ for a CNOT with control $i$ and target $j$.
For $n$ qubits, define the two-layer brickwork CNOT circuit
\begin{align}
    U_n &= U_{\mathrm{even}}\, U_{\mathrm{odd}},
    \nonumber\\
    U_{\mathrm{odd}}
    &=
    \prod_{i=1}^{\lfloor n/2\rfloor}
    \mathrm{CNOT}_{2i-1\to 2i},
    \nonumber\\
    U_{\mathrm{even}}
    &=
    \prod_{i=1}^{\lfloor (n-1)/2\rfloor}
    \mathrm{CNOT}_{2i\to 2i+1},
    \label{eq:brickwork-Un}
\end{align}
where $U_{\mathrm{odd}}$ is applied first and the CNOTs within each layer are mutually disjoint.
Hence $U_n$ has constant circuit depth for all $n$.
As shown in Ref.~\cite{xu2026controlled}, this family has high Clifford periodicity: setting
\begin{equation}
    m = \lceil \log_2 n \rceil,
    \label{eq:m-definition}
\end{equation}
the unitary $U_n$ has period $2^m$, and its eigenvalues are $2^m$-th roots of unity.
Let $\omega_m = e^{2\pi\ii/2^m}$ denote the primitive $2^m$-th root of unity.
For the orbit generated from the computational-basis state $\ket{10\cdots 0}$, define
\begin{equation}
    \ket{\psi_{n,r}}
    =
    \frac{1}{\sqrt{\mathcal N_n}}
    \sum_{k=0}^{2^m-1}
    \omega_m^{-rk}\, U_n^k \ket{10\cdots 0},
    \label{eq:psi-n-r}
\end{equation}
where $\mathcal N_n$ is the normalization.
For the brickwork instances used below the orbit has length $2^m$, so $\ket{\psi_{n,r}}$ is an equal-weight superposition over $2^m$ computational-basis states.
By construction, $\ket{\psi_{n,r}}$ is an eigenstate of $U_n$ with eigenvalue $\omega_m^r$:
\begin{equation}
    U_n \ket{\psi_{n,r}} = \omega_m^r \ket{\psi_{n,r}}.
    \label{eq:Un-eigenstate}
\end{equation}
An ideal controlled-$U_n$ gate therefore implements phase kickback onto the control qubit:
\begin{align}
    C U_n
    \left[
        \left(\alpha\ket{0}+\beta\ket{1}\right)
        \otimes \ket{\psi_{n,r}}
    \right]
    =
    \left(\alpha\ket{0}+\beta\, \omega_m^r\ket{1}\right)
    \otimes \ket{\psi_{n,r}}
    \nonumber\\
    =
    \left( Z^{r/2^{m-1}} \otimes I \right)
    \left[
        \left(\alpha\ket{0}+\beta\ket{1}\right)
        \otimes \ket{\psi_{n,r}}
    \right].
    \label{eq:phase-kickback}
\end{align}
Choosing $r = 1$ and $m = b+1$ therefore yields an exact dyadic phase $Z^{2^{-b}}$, returning the catalyst to itself.
The smallest brickwork instance with $m = b+1$ has $n = 2^b + 1$, so this direct construction uses $O(2^b)$ logical catalyst qubits.
 
Three properties of this construction make it well-suited to surface-code cultivation.
First, the catalyst is an eigenstate of a \emph{Clifford} circuit, so its verification can be organized around controlled-Clifford checks rather than the non-Clifford measurements that would be required to directly verify a dyadic magic state with $b \geq 3$. In fact, $U_n$ has a set of degenerate eigenstates for $r=1$. This degeneracy is harmless for the catalytic application because only the eigenphase matters.
Second, $U_n$ contains only CNOT gates, whose logical implementation between corresponding blocks of a CSS code is always transversal.
Third, the controlled version of each CNOT in $U_n$ is a three-qubit Toffoli gate, which connects directly to the standard non-Clifford resources used in modern fault-tolerant compilations: $\ket{T}$, $\ket{\mathrm{CCZ}}$, and measurement-assisted logical-AND constructions~\cite{jones2013low,selinger2013quantum,gidney2018halving}.

A further advantage is that the online invocation of the catalyst can be made \emph{constant} in $T$-depth: by fanning the phase-estimation control register out through a depth-$O(b)$ CNOT tree, the Toffoli gates that implement $CU_n$ can be applied in parallel, in close analogy with the constant-$T$-depth construction of Ref.~\cite{kim2025catalytic}.
We note that the brickwork family in Eq.~\eqref{eq:brickwork-Un} is not the most qubit-efficient high-period Clifford construction in Ref.~\cite{xu2026controlled}, but its CNOT-only structure makes both the online invocation circuit and the logical verification circuit structurally simple, which is what we now exploit.
 
\subsection{The \texorpdfstring{$\sqrt{T}$}{sqrt(T)} catalyst}
\label{sec:protocol-sqrtT}
 
In the rest of our work, we focus on $n = 9$.
Then $m = \lceil \log_2 9\rceil = 4$, so $U_9$ has period $16$ and its eigenvalues are $16$th roots of unity.
The brickwork circuit is
\begin{align}
    U_9 &=
    \big(\mathrm{CNOT}_{2\to3}\,
    \mathrm{CNOT}_{4\to5}\,
    \mathrm{CNOT}_{6\to7}\,
    \mathrm{CNOT}_{8\to9}\big)
    \nonumber\\
    &\quad\times
    \big(\mathrm{CNOT}_{1\to2}\,
    \mathrm{CNOT}_{3\to4}\,
    \mathrm{CNOT}_{5\to6}\,
    \mathrm{CNOT}_{7\to8}\big),
    \label{eq:U9-circuit}
\end{align}
with the second factor (the odd layer) applied first.
The target catalyst is the $r = 1$ eigenstate, with eigenvalue $e^{\ii\pi/8}$. In fact, there are in total 16 degenerate eigenstates with such a phase. One representative is 
\begin{align}
    &\ket{\psi_9}
    =
    \frac{1}{4}
    \sum_{k=0}^{15}
    e^{-\ii\pi k/8}\, U_9^k\ket{100000000},
    \nonumber\\
    &U_9\ket{\psi_9} = e^{\ii\pi/8}\ket{\psi_9}.
    \label{eq:psi9}
\end{align}
An ideal controlled-$U_9$ gate therefore applies
$Z^{1/8} = \operatorname{diag}(1, e^{\ii\pi/8}) = \sqrt{T}$
to the control qubit, returning $\ket{\psi_9}$ to itself. Other degenerate eigenstates with the same phase can be similarly constructed.

The online invocation cost is governed by the structure of $CU_9$.
Since $U_9$ contains eight CNOT gates, $CU_9$ contains eight controlled-CNOTs, i.e.\ eight Toffoli-type gates.
A conservative unitary Clifford+$T$ implementation costs $8 \times 7 = 56$ $T$ gates per invocation.
If measurement and feedforward are allowed, each Toffoli can be replaced by a measurement-assisted logical-AND~\cite{gidney2018halving}, in which a temporary logical AND is computed using 4 $T$ gates, a Clifford CNOT copies it into the target, and the temporary register is uncomputed by measurement with no further $T$ gates. This reduces the cost to $32$ $T$ gates.
Alternatively, we can use 8 $\ket{\mathrm{CCZ}}$ states~\cite{eastin2013distilling,jones2013low,haah2018codes,gidney2019efficient} from distillation that can be consumed in parallel.
In every variant, the magic-state cost is constant in the fineness of the target dyadic phase, and no rotation-synthesis approximation error is introduced when the catalyst is invoked.
 
\subsection{The cultivation protocol}
\label{sec:protocol-cultivation}
 
Our cultivation protocol prepares a fault-tolerant logical copy of $\ket{\psi_9}$ in three stages.
A physical $\ket{\psi_9}$ is first prepared and encoded into nine independent rotated-surface-code blocks of distance three.
The encoded state is then verified by a logical-$U_9$ phase estimation measurement and screened by distance-three stabilizer measurement.
Finally, the postselected blocks are grown to distance seven through a sequence of unitary code growth, stabilizer-measurement growth, and complementary-gap decoding.
Each stage uses only Clifford operations on the encoded state, so the entire verification and growth pipeline can be performed by standard fault-tolerant gadgets.
The detailed noise model and simulation methodology are deferred to Sec.~\ref{sec:simulation}.
 
Throughout the remainder of the paper, we denote the rotated and regular (unrotated) surface codes of distance $d$ as $\rm Rot(d)$ and $\rm Reg(d)$, respectively.

\subsubsection{Physical preparation and encoding}
\label{sec:protocol-cultivation-encoding}
 
The cultivation begins with a physical nine-qubit state on the $\ket{\psi_9}$ orbit, which can be initialized by any of several methods consistent with the protocol's noise budget, such as a direct unitary synthesis or a physical quantum phase estimation (QPE). 
The nine physical qubits are then encoded into nine independent distance-three rotated surface code blocks using a unitary encoding circuit $U_{\rm enc}^{\otimes9}$. Details of this circuit are given in Appendix~\ref{app:scencode}. 
 
After encoding, the logical state on the nine blocks is nominally $\ket{\overline{\psi_9}}$, but it inherits both the noise of the physical preparation and the noise of the encoding gates.
The role of the subsequent verification and growth stages is to detect catalyst errors and project onto a high-fidelity logical $\ket{\psi_9}$.
 
\subsubsection{Logical-$U_9$ verification by phase estimation}
\label{sec:protocol-cultivation-verification}
 
\begin{figure*}[t]
    \centering
    \includegraphics[width=0.8\linewidth]{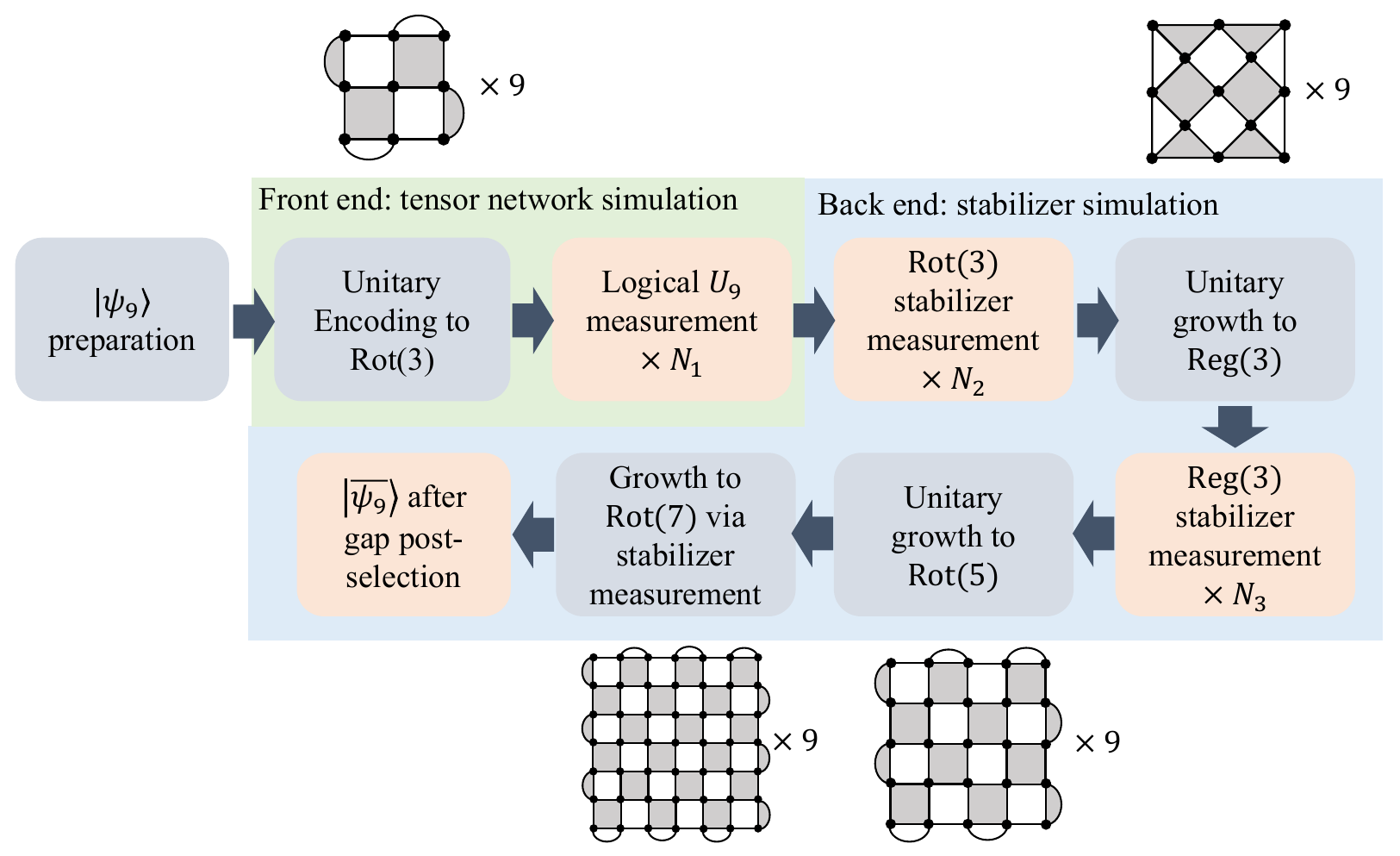}
    \caption{The cultivation pipeline for logical catalyst state $\ket{\overline{\psi_9}}$. The steps marked in orange involve postselection. The front/back end division in our numerical simulation of the protocol is marked in the figure. }
    \label{fig:pipeline}
\end{figure*}
 
The central verification step is a logical-$U_9$ measurement implemented by a logical-level QPE circuit.
Since the target eigenvalue $e^{\ii\pi/8} = e^{2\pi\ii/16}$ corresponds to one of sixteen possible eigenphases, four ancilla bits suffice to resolve it.
A verification round uses four fresh ancilla qubits $a_1, a_2, a_3, a_4$, each initialized in $\ket{+}$, and applies the standard QPE block
\begin{equation}
    \prod_{j=1}^{4}
    C_{a_j}\!\left(\bar U_9^{\,2^{4-j}}\right),
    \label{eq:qpe-controlled-powers}
\end{equation}
in which $a_j$ controls the logical Clifford power $\bar U_9^{2^{4-j}}$.
 
Since each logical CNOT in $\bar U_9$ is transversal between the corresponding $\mathrm{Rot}(3)$ blocks, a controlled logical CNOT decomposes into nine physical Toffoli gates.
As $U_9$ contains eight logical CNOTs, a direct sequential implementation of $C\bar U_9$ would contain $8 \times 9 = 72$ physical Toffoli gates, all sharing the same phase-estimation ancilla.
To increase parallelism, we fan the ancilla out into a length-three GHZ state and use the three GHZ qubits to control three lanes of the transversal operation, reducing the per-block depth from nine Toffoli gates to three.
After the controlled operation the GHZ state is uncomputed, and its two non-readout qubits are measured and postselected on the trivial cleanup outcome.
This GHZ fan-out strategy is already used for logical measurements in fold-transversal surface-code cultivation~\cite{sahay2025fold}, where it is shown to be consistent with the fault distance three of the distance-three surface code.
 
The higher powers $U_9^2$, $U_9^4$, and $U_9^8$ are also Clifford circuits, and we implement them using compressed CNOT schedules rather than by literal repeated application of $U_9$.
The schedules used in our circuit have CNOT counts
\begin{equation}
    \#\mathrm{CNOT}(U_9, U_9^2, U_9^4, U_9^8) = (8, 15, 10, 1),
    \label{eq:power-cnot-counts}
\end{equation}
with $U_9^8 = \mathrm{CNOT}_{1\to 9}$.
Explicit CNOT decompositions of $U_9^2$ and $U_9^4$ are given in Appendix~\ref{app:explicitcnots}.
 
After the controlled powers, a conventional QPE round would apply an inverse four-qubit quantum Fourier transform~\cite{kitaev2002classical}.
Since the ancillas are measured immediately and only the outcome corresponding to eigenvalue $e^{\ii\pi/8}$ is accepted, we instead use the semiclassical inverse QFT of Griffiths and Niu~\cite{griffiths1996semi} with the phase corrections fixed to the values consistent with the accepted eigenphase, see the circuit diagram in Fig.~\ref{fig:logicalu9circ}.
This avoids both a coherent inverse-QFT block and feedback circuit on the ancilla register without changing the accepted branch.
We postselect runs that produce the accepted phase-estimation outcome\footnote{In the conventional QFT circuit the readout registers are bit-reversed at the end; the binary fraction $0.0001_2$ for $e^{\ii\pi/8}$ therefore corresponds to the ancilla readout string $1000$.}
\begin{equation}
    1000.
    \label{eq:target-logical-bits}
\end{equation}
In an actual cultivation protocol, the logical $U_9$ measurement can be repeated multiple times to increase the logical fidelity of the catalyst state.  We denote the number of times $U_9$ is measured by $N_1$. We benchmark the result for different numbers of rounds of logical $U_9$ measurement in Sec.~\ref{sec:simulation}.
 
Although we initialize the logical state as $\ket{\psi_9}$, the eigenvalue $e^{\frac{\ii\pi}{8}}$ is $16$-fold degenerate, so noise may rotate the accepted state into a different eigenvector that shares this eigenphase.
For use as a catalyst this is harmless. Only the eigenphase is kicked back, not the particular eigenvector, so the sole damaging error is \emph{leakage} into an eigenspace with a different eigenphase.
The phase-estimation read-out is an \emph{exact} check against such leakage in the following sense: since $U_9$ has period $16$, its eigenvalues lie precisely in the set $\{e^{2\pi\ii m/16}\}_{m=0}^{15}$. Hence, the phase estimation has no spectral leakage, and the four-bit string is a faithful logical syndrome of the eigenphase.
Accepting $1000$ therefore projects the state \emph{exactly} into the $e^{\frac{\ii\pi}{8}}$ eigenspace, putting leakage and a nontrivial read-out in one-to-one correspondence: without measurement error no single physical fault can leak the eigenphase undetected.
 
Crucially, a single verification round ($N_1=1$) is already enough, thanks to the inherent fault-tolerance of the logical $U_9$ eigenspace.
To see this, we propagate every single-qubit Pauli fault on the physical state $\ket{\psi_9}$ through an ideal phase estimation, the details of which are presented in Appendix~\ref{app:u9ft}. We find that the only faults escaping detection act as a global phase and are harmless, while a detected fault produces a read-out whose Hamming distance from $1000$ averages $\bar w = 1.654$.
Because masking a leak requires flipping these read-out bits with additional measurement faults, a single logical-$U_9$ measurement carries an effective logical fault distance
\begin{equation}
      \mathbf{f}_{\mathrm{logical}} =  1 + \bar w \approx 2.654,
  \label{eq:f-logical}
\end{equation}
where the ``$1$'' counts the physical fault and $\bar w$ the additional read-out faults needed to forge $1000$. This means a \emph{single} logical-$U_9$ round already exceeds fault distance two.
As a result, once the growth stage supplies a stabilizer fault distance $\mathbf{f}_{\mathrm{stab}}=3$, the decoded logical leakage rate would scale as $O(p^{\lceil\frac{\min \{\mathbf{f}_{\mathrm{stab}},\mathbf{f}_{\mathrm{logical}} \}}{2}\rceil})=O(p^2)$, and can be suppressed further by complementary-gap postselection, as we confirm in the numerical simulation in Sec.~\ref{sec:simulation}. This is in stark contrast to the cultivation of $\ket{T}$~\cite{gidney2024magic,sahay2025fold,claes2025cultivating,chen2025rp2}, where the logical check is done twice to reach fault distance three. 
 
\subsubsection{Stabilizer measurement, code growth, and escape to $\mathrm{Rot}(7)$}
\label{sec:protocol-cultivation-growth}
 
A logical-$U_9$ verification round checks the catalyst eigenvalue, but does not by itself guarantee that the nine code blocks remain within the code space.
We therefore follow the logical-$U_9$ check and ancilla cleanup with noisy distance-three stabilizer-measurement rounds on every $\mathrm{Rot}(3)$ block.
Only shots in which every observed stabilizer measurement is zero are passed to the growth stage. This hard postselection removes the leading detectable error events from the surviving population before any code-distance enlargement. In practice, we can repeat the stabilizer measurement and postselection for $N_2$ times, which could potentially increase the logical quality.
 
The accepted state is then grown blockwise.
The first growth step is a unitary rotation $\mathrm{Rot}(3) \longrightarrow \mathrm{Reg}(3)$, which introduces four additional data qubits per block.
We then measure all twelve $\mathrm{Reg}(3)$ stabilizers and postselect on the trivial outcome. In practice, this can be repeated for $N_3$ times.
After the $\mathrm{Reg}(3)$ check(s), each block is grown unitarily to $\mathrm{Rot}(5)$.
Finally, the block is enlarged from $\mathrm{Rot}(5)$ to $\mathrm{Rot}(7)$ by stabilizer-measurement growth, following the approach outlined in Ref.~\cite{Li2015a} and the surface-code growth/escape structure used in recent cultivation protocols~\cite{gidney2024magic,claes2025cultivating,sahay2025fold}.
 
Some measurements during this final growth to $\rm Rot(7)$ surface codes correspond to newly introduced degrees of freedom and have intrinsically random outcomes. These measurements are not used as hard postselection checks.
Instead, the full space-time syndrome record is decoded with a matching-based soft decoder, and final acceptance is controlled by the complementary gap between the best and next-best logical Pauli-frame hypotheses.
In our implementation each of the nine blocks contributes two logical-frame bits, corresponding to residual logical-$X$ and logical-$Z$ frame components, so the full cultivated catalyst carries an 18-bit decoded Pauli frame.
 
The entire cultivation pipeline is summarized in Fig.~\ref{fig:pipeline}.
 
\section{Simulation protocol and results}\label{sec:simulation}
 
\subsection{Hybrid simulation strategy}
\label{sec:simulation-strategy}
 
We estimate the performance of the cultivation circuit of
Sec.~\ref{sec:protocol-cultivation} by a shot-by-shot simulation of the
full noisy circuit. The front end of the protocol, i.e. the physical
preparation of $\ket{\psi_9}$ and the logical-$U_9$ verification, contains
a large number of non-Clifford physical gates: the Toffoli gates that
implement the controlled powers of $\bar U_9$, and the $S^\dagger$,
$T^\dagger$, $\sqrt{T}^\dagger$ corrections of the semiclassical inverse
QFT, see Fig.~\ref{fig:logicalu9circ}. Therefore, it cannot be handled by a
stabilizer simulator. The growth stage, by contrast, is Clifford. We split
the simulation at the end of the logical-$U_9$ verification into a
non-Clifford \emph{front end}, evolved as a state vector, and a Clifford
\emph{back end}, evolved as a stabilizer tableau. This is inspired by the handoff
strategy of Ref.~\cite{sahay2025fold}: for stochastic Pauli noise it
reproduces the statistics of the full physical circuit exactly, because
measuring the code syndrome at the split point collapses the front-end
state to a definite syndrome together with a logical state that is carried
forward intact; see Eq.~\eqref{eq:handoff-decomp}.
 
Throughout the simulation, we use a circuit-level depolarizing model at physical
error rate $p$, realized stochastically as Pauli errors so that the front
end can be evolved as a pure state in each shot. Every $a$-qubit gate
($a=1,2,3$) is followed by the depolarizing channel
\begin{equation}
  \mathcal{E}_a(\rho)
  = (1-p)\,\rho
  + \frac{p}{4^{a}-1}
    \sum_{\substack{P\in\mathsf{P}_a\\ P\neq\id}} P\rho P^\dagger ,
  \label{eq:depol-channel}
\end{equation}
where $\mathsf{P}_a$ is the $a$-qubit Pauli group.
Each single-qubit initialization is faulty with probability $p$ (the state
is replaced by the orthogonal state in its preparation basis), and each
measurement outcome is flipped with probability $p$. We take
$p=10^{-3}$ unless otherwise stated.
 
\subsubsection{Front end: tensor-network state vector}
The first simulator carries
the protocol from the preparation of the physical catalyst $\ket{\psi_9}$
through encoding by $\Uenc^{\otimes 9}$ and the logical-$U_9$ verification
round(s), up to and including the ancilla measurements: the four
phase-estimation read-outs $a_{j,1}$ and the GHZ clean-up qubits
$a_{j,2},a_{j,3}$ ($j=1,\dots,4$). The $81$-qubit data register $\Dcal$ and
the ancillas are evolved as a matrix-product state (MPS) using
\texttt{iTensor}~\cite{itensor}. We find that the MPS for all the qubits in the protocol before measurement (93 for one round of $U_9$ measurement and 105 for two rounds of $U_9$ measurement) converges with bond dimension 256 (see Appendix~\ref{app:tnconverge}), which will be used in the subsequent simulation. 
 
In each shot the gate errors of
Eq.~\eqref{eq:depol-channel} are sampled as concrete Pauli operators, and a
measurement error is realized by inserting, immediately before each
measurement, the Pauli anticommuting with the measured observable with
probability $p$ ($Z$-type before the $X$-basis read-outs $a_{j,1}$,
$X$-type before the $Z$-basis clean-up read-outs).
Rather than drawing the ancilla outcomes stochastically and discarding the
shots that miss the target, we evaluate the front-end postselection
\emph{analytically} for each fault sample. Let $\ket{\Psi_i}$ be the
normalized front-end state of shot $i$ after the sampled gate- and
measurement-error Paulis have been applied, immediately before the ancilla
read-outs, and let
\begin{equation}
  \Pi_{\mathrm{acc}}=\Pi^{\mathrm{PE}}_{1000}\otimes\Pi^{\mathrm{GHZ}}_{0}
  \label{eq:accept-projector}
\end{equation}
be the projector onto the accepted phase-estimation string $1000$ on the
four read-out ancillas $a_{j,1}$ (cf.\ Eq.~\eqref{eq:target-logical-bits})
and the trivial outcome on the GHZ clean-up qubits $a_{j,2},a_{j,3}$. For
each shot $i$, we record the \emph{Born weight}
\begin{equation}
  w_i=\bigl\|\,\Pi_{\mathrm{acc}}\ket{\Psi_i}\bigr\|^{2}
     =\bra{\Psi_i}\Pi_{\mathrm{acc}}\ket{\Psi_i},
  \label{eq:born-weight}
\end{equation}
the probability that a literal stochastic run of this fault sample would
pass the front-end postselection, and carry forward the
normalized post-measurement state
$\ket{\Psi_i^{\mathrm{acc}}}=\Pi_{\mathrm{acc}}\ket{\Psi_i}/\sqrt{w_i}$ whenever $w_i>0$. This state is a pure, generally non-stabilizer, nine-block logical state that is passed to
the handoff step below. Because $\Pi_{\mathrm{acc}}$ marginalizes the ancilla
outcomes exactly, this procedure retains \emph{every} sampled shot except the
measure-zero set with $w_i=0$ (fault configurations for which the target
outcome is unreachable). Each retained shot then enters all subsequent
averages with statistical weight $w_i$. This is a Rao--Blackwellization~\cite{rao1945information,blackwell1947conditional,robert2004monte} of the
postselection: it removes the sampling noise of the accept/reject step and
leaves essentially all $N_f$ front-end shots available to seed the back end,
at the price of carrying the weights $w_i$.
 
\subsubsection{Handoff}
To transfer the accepted front-end state to the stabilizer
simulator we read out the code syndrome \emph{noiselessly}. Applying
$(\Uenc^{\dagger})^{\otimes 9}$ maps each $\mathrm{Rot}(3)$ block back to one
bare logical qubit and eight disentangled syndrome qubits. We then measure
the $9\times 8=72$ syndrome qubits.
We stress that this read-out should carry NO noise: it only extracts the true
syndrome eigenvalues of the front-end state to seed the back end. The physically noisy
distance-three syndrome round is the \emph{first} round of the back end.
 
At this point, the accepted front-end state has the form
\begin{equation}
  \ket{\Psi_{\rm front}}
  = \sum_{s\in\mathbb{F}_2^{72}} c_s\,
    \ket{s}_{\rm syn}\otimes\ket{\bar\phi_s}_{\rm log},
  \label{eq:handoff-decomp}
\end{equation}
so the syndrome measurement returns a particular $s$ with probability
$|c_s|^2$ and collapses the nine logical qubits to
$\ket{\bar\phi}\equiv\ket{\bar\phi_s}$. We record $s$, with per-block
components $s_b\in\mathbb{F}_2^{8}$, and store the nine-qubit logical state
vector $\ket{\bar\phi}$ for the final leakage evaluation. Because the front
end contains non-Clifford gates, the logical error carried by $\ket{\bar\phi}$
need not be a Pauli operator; retaining the full logical state vector,
rather than a Pauli frame, is exactly what makes the handoff faithful for
the magic state.
 
\subsubsection{Back end: stabilizer simulation of growth}
The remaining
stages, including the noisy $\mathrm{Rot}(3)$ syndrome round, the unitary growth
$\mathrm{Rot}(3)\!\to\!\mathrm{Reg}(3)$ with its syndrome round, and the
growth $\mathrm{Reg}(3)\!\to\!\mathrm{Rot}(5)\!\to\!\mathrm{Rot}(7)$, are
Clifford and are simulated with Stim~\cite{gidney_stim_2021}. Their sole
purpose is to determine the residual logical Pauli error left on the
catalyst after decoding. Two facts let us simulate the nine blocks
\emph{independently}: the growth applies no entangling gates between blocks,
and the residual logical Pauli error of a stabilizer circuit is independent
of the encoded logical state. Consequently, each block $b$ can be initialized by applying
$\Uenc$ to $\ket{\bar 0}$ on its logical qubit and to $\ket{s_b}$ on its
syndrome qubits, so that the back-end state carries the handoff syndrome
$s_b$ of that block.
 
In each $\mathrm{Rot}(3)$ and $\mathrm{Reg}(3)$ syndrome round we hard-postselect
shots whose measured stabilizer values are all trivial, jointly
across the nine blocks.
The final enlargement $\mathrm{Rot}(5)\!\to\!\mathrm{Rot}(7)$ introduces
measurements of newly created degrees of freedom whose first outcomes are
intrinsically random; these are not used as hard checks. Instead the full
space-time detector record of each block is decoded with PyMatching~\cite{higgott2022pymatching}, and final acceptance is controlled
by the complementary gap between the two logical-coset hypotheses, evaluated
per block and per logical observable. A shot is accepted only if every block
clears the gap threshold.
 
\subsubsection{Residual logical leakage}
For each accepted shot and each
block $b$, Stim provides the logical-$\bar X_b$ and logical-$\bar Z_b$ frame
flips \emph{actually} produced by the sampled physical errors, while the
decoder provides its \emph{prediction} of the same two bits. The residual
logical error is the bit-wise XOR of the two,
\begin{equation}
  (x_b,z_b)
  = \bigl(
      \ell^{X,\mathrm{act}}_b \oplus \ell^{X,\mathrm{pred}}_b,\;
      \ell^{Z,\mathrm{act}}_b \oplus \ell^{Z,\mathrm{pred}}_b
    \bigr),
  \label{eq:residual-frame}
\end{equation}
which vanishes when the decoder succeeds on block $b$ and equals the
uncorrected logical Pauli otherwise.
Collecting the bits into $x,z\in\mathbb{F}_2^{9}$, the back-end errors act on
the stored logical state as $\prod_{b=1}^{9}\bar X_b^{\,x_b}\bar Z_b^{\,z_b}$. As argued at the end of Sec.~\ref{sec:protocol-cultivation-verification}, for the catalyst state to be fully functional, one has to make sure that the error-corrected state lands in the 16-fold degenerate eigen-subspace of $U_9$ with eigenvalue $e^{\frac{\ii\pi}{8}}$. Therefore, we define the logical fidelity as
\begin{equation}
  F
  = \Bigl|
      P_{r=1}\,
      \prod_{b=1}^{9}\bar X_b^{\,x_b}\bar Z_b^{\,z_b}\,
      \ket{\bar\phi}
    \Bigr|^{2},
  \label{eq:shot-fidelity}
\end{equation}
where $P_{r=1}$ is the projector onto the $e^{\frac{\ii\pi}{8}}$ phase eigenspace. Hence, we can further define the \textit{logical leakage rate} $L\equiv 1-F$ to characterize the infidelity of the cultivated state when used in catalytic logical phase gates.
 
\begin{figure*}[t]
    \includegraphics[width=0.6\linewidth]{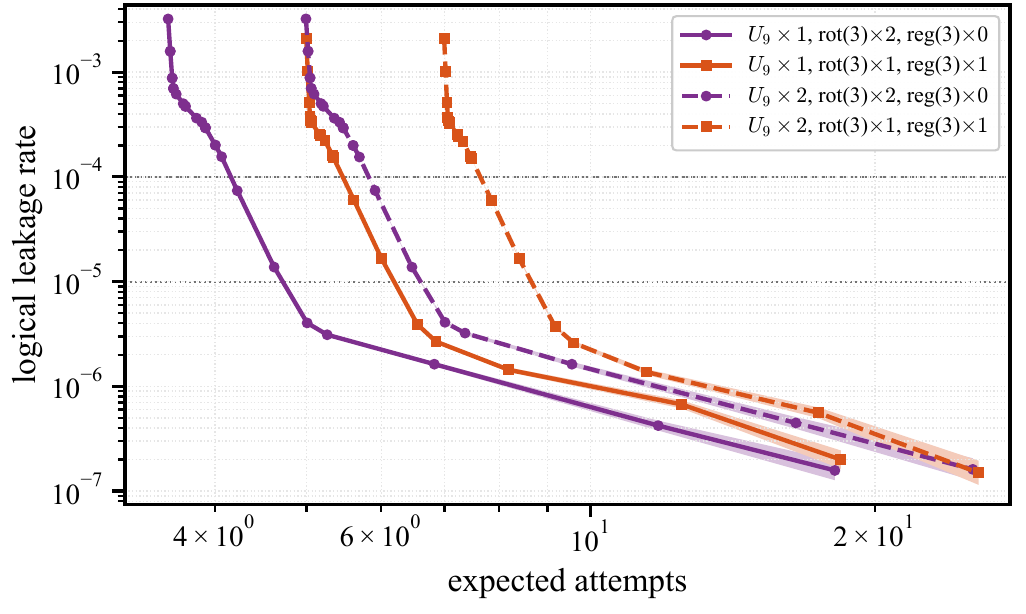}
    \caption{Simulation result of the cultivation protocol for 
    $U_9\times N_1,\mathrm{Rot}(3)\times N_2,\mathrm{Reg}(3)\times N_3$, with $N_1=1,2$ and $(N_2,N_3)=(1,1)$ and $(2,0)$. Each curve is obtained by sweeping the decoder complementary-gap threshold from $0$ to $60$ at physical error rate $p=10^{-3}$.}
    \label{fig:cultivation-comparison}
\end{figure*}

Once the simulation is done, we need to pool across all the shots. Each front-end shot $i$ ($i=1,\dots,N_f$) carries the Born weight $w_i$
of Eq.~\eqref{eq:born-weight} and seeds a set of back-end growth samples
conditioned on its handoff syndromes $s^{(i)}_b$. For a complementary-gap
threshold $g$, let $\alpha_i(g)\in[0,1]$ be the fraction of shot $i$'s
back-end samples that clear all $\mathrm{Rot}(3)$ and $\mathrm{Reg}(3)$
stabilizer checks and the gap test, and let $L_i(g)$ be the mean per-shot
leakage $L=1-F$, computed using Eq.~\eqref{eq:shot-fidelity}, over those accepted samples
for which $\alpha_i(g)>0$. Weighting every shot by its Born weight,
the reported logical leakage rate and expected number of attempts are
\begin{align}
  \bar L(g) &= \frac{\sum_{i=1}^{N_f} w_i\,\alpha_i(g)\,L_i(g)}
                    {\sum_{i=1}^{N_f} w_i\,\alpha_i(g)},
  \label{eq:leakage-estimator}\\[4pt]
  \overline{N_{\mathrm{att}}}(g)
            &= \Bigl[\,\tfrac{1}{N_f}\textstyle\sum_{i=1}^{N_f}
               w_i\,\alpha_i(g)\,\Bigr]^{-1}.
  \label{eq:attempts-estimator}
\end{align}
The bracketed quantity in Eq.~\eqref{eq:attempts-estimator} is the overall
acceptance probability of a single attempt; it factorizes into the front-end
logical-$U_9$ survival $\tfrac{1}{N_f}\sum_i w_i$ and the conditional
back-end survival, i.e. the hard $\mathrm{Rot}(3)$/$\mathrm{Reg}(3)$ syndrome
checks together with the gap acceptance as carried by $\alpha_i(g)$. Sweeping
$g$ traces the leakage--attempts trade-off curves of
Fig.~\ref{fig:cultivation-comparison}.

\subsection{Simulation result}
 
We now present results for the full cultivation protocol of
Sec.~\ref{sec:protocol-cultivation}, which admits a family of strategies set
by how much postselection is applied at each stage. In the front end the
logical-$U_9$ verification may be repeated $N_1$ times, accepting only shots
whose every four-bit phase-estimation read-out equals the target string $1000$
[Eq.~\eqref{eq:target-logical-bits}] and whose GHZ clean-up qubits all return
the trivial outcome. In the back end we may insert $N_2$ rounds of
$\mathrm{Rot}(3)$ and $N_3$ rounds of $\mathrm{Reg}(3)$ syndrome extraction,
each hard-postselected on an all-trivial \emph{raw} syndrome across the nine
blocks. We label a strategy by
$U_9\times N_1,\mathrm{Rot}(3)\times N_2,\mathrm{Reg}(3)\times N_3$. Every
strategy terminates identically, with unitary growth to $\mathrm{Rot}(5)$, a
stabilizer round enlarging to $\mathrm{Rot}(7)$, and a final per-block
acceptance on the complementary gap of the matching decoder. For every protocol, we generate $2000$ shots from the front-end tensor-network state vectors, and those that pass postselection undergo a back-end simulation with $2\times10^{6}$ samples.
Fig.~\ref{fig:cultivation-comparison} compares four such strategies
($N_1\in\{1,2\}$, $(N_2,N_3)\in\{(1,1),(2,0)\}$) with the fewest attempts. 
 
Each curve in Fig.~\ref{fig:cultivation-comparison} is traced by sweeping the complementary-gap threshold: as the threshold is raised, more shots are rejected (so the expected number of attempts grows), while the surviving population is cleaner, so the leakage falls.
At the loosest threshold, every route starts near a logical leakage rate of $3\times10^{-3}$, and all four converge to $\sim10^{-7}$ once the threshold is tightened.
Two features stand out.
First, $\mathrm{Rot}(3)$ postselection is slightly more economical than $\mathrm{Reg}(3)$: at equal leakage, $(N_2,N_3)=(2,0)$ reaches the target with fewer attempts than $(1,1)$.
Second, both routes shown use $N_2+N_3=2$ rounds of distance-three checks, which is what supplies the stabilizer fault distance $\mathbf{f}_{\mathrm{stab}}=3$, needed to suppress the leading single-qubit errors. The acceptance and gap-threshold dependence, and the routes with fewer or more growth rounds, are compared in full in Appendix~\ref{app:growth}.
Figure~\ref{fig:discardpie} shows where the rejected shots are discarded across the postselection stages for the two $N_1=1$ routes.
With a single logical-$U_9$ round, either route attains a logical leakage rate $O(p^2)\sim10^{-6}$ at roughly $7$ to $8$ expected attempts.
 
\begin{figure}[h]
    \centering
    \includegraphics[width=0.9\linewidth]{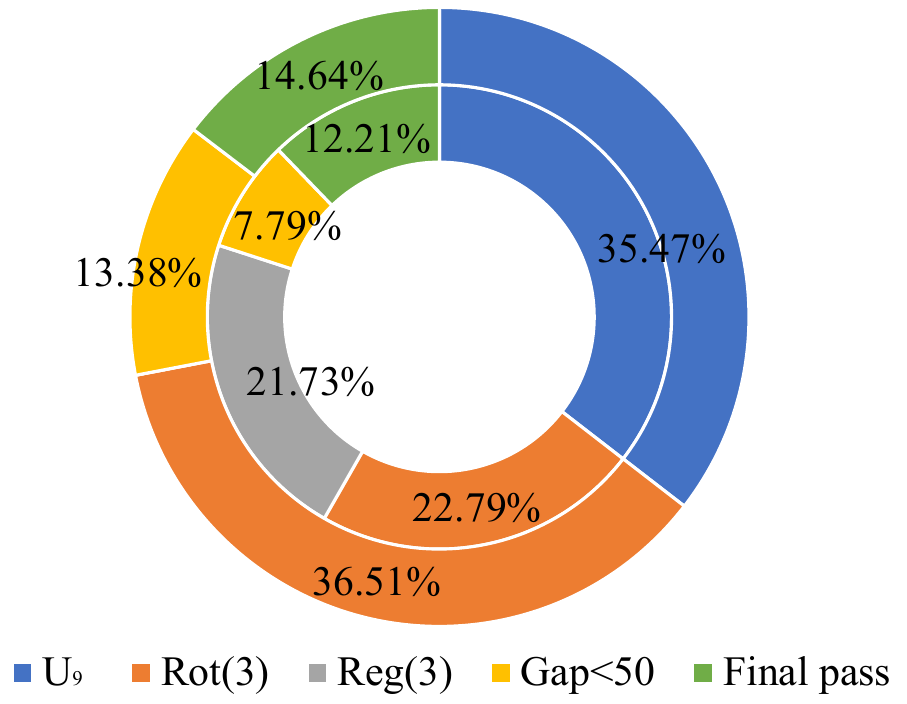}
    \caption{Percentage of shots that are discarded in each stage of postselection for $(N_1,N_2,N_3)=(1,1,1)$ (inner) and $(N_1,N_2,N_3)=(1,2,0)$ (outer). The complementary gap in this case is fixed to be 50dB.}
    \label{fig:discardpie}
\end{figure}
 
The most salient feature of Fig.~\ref{fig:cultivation-comparison} is that the single-round ($N_1=1$) routes match the double-round ($N_1=2$) routes in leakage while using far fewer attempts.
This is the numerical signature of the effective logical fault distance $\mathbf{f}_{\mathrm{logical}}\approx2.654$ of Eq.~\eqref{eq:f-logical}: because one logical-$U_9$ read-out already exceeds fault distance two, a second round is not needed to reach the leading $O(p^{2})$ scaling, and the curves descend from this $\sim10^{-6}$ level toward the $O(p^{\mathbf{f}_{\mathrm{logical}}})=O(p^{2.654})\sim10^{-8}$ undetectable-fault floor as the gap threshold is tightened.
Doubling the verification of logical $U_9$ ($N_1=2$) only lowers this deep floor while leaving the leading scaling unchanged. In fact, the marginal benefit of the extra check is essentially canceled by the additional errors its QPE circuit introduces, so it buys cleaner catalysts only at a substantially higher attempt cost.
 
Finally, we stress that the noise model used throughout is deliberately crude.
The circuit-level depolarizing model of Eq.~\eqref{eq:depol-channel} assigns the \emph{same} error rate $p$ to every physical operation, including one-, two-, and three-qubit gates, state preparations, and measurements alike. It neglects idling errors entirely, so qubits that wait while other operations are carried out accumulate no noise.
A realistic device hosts much more complicated noise profile: entangling gates are typically noisier than single-qubit gates, multi-qubit gates are noisier if carried over geometrically non-local qubits, and idling over the considerable depth of the logical-$U_9$ verification adds further dephasing.
Such a hardware-calibrated noise model can be inserted directly into the same hybrid simulation, without any change to the methodology.
However, we expect our conclusions to be robust: because the protocol already reaches a logical leakage rate of $\sim10^{-7}$ in the regime studied, restoring a realistic level of noise should raise this by at most roughly an order of magnitude, to $\sim10^{-6}$.
This remains comparable to the logical infidelity reported for existing cultivation protocols that prepare a single-block magic state such as $\ket{T}$~\cite{gidney2024magic,chen2025rp2,sahay2025fold,claes2025cultivating}, even though our catalyst supports a finer dyadic phase across nine surface-code blocks rather than one.

\section{Resource comparison}\label{sec:application}

Having established the construction and its cultivation, we now position the $\sqrt{T}$ catalyst, and the general catalyst states for the $Z^{2^{-b}}$ gate family, against the standard routes to fine dyadic phase rotations. A fair comparison must keep two costs distinct: the \emph{one-time preparation} cost of any resource state, which is amortized over all subsequent uses, and the \emph{online} cost paid on each invocation of the gate. Table~\ref{tab:comparison} collects both. The four methods fall into two families. Per-gate synthesis, which includes ancilla-free Clifford+$T$ synthesis~\cite{dawson2006solovay,kliuchnikov2013fast,ross2016optimal} and repeat-until-success (RUS) circuits~\cite{paetznick2014repeat,bocharov2015efficient}, carries no resource state but repays an approximation cost on \emph{every} use. Catalytic methods, including the conventional phase-gradient state~\cite{kitaev1995quantum,kitaev2002classical,gidney2018halving} and our construction, prepare a reusable state once and apply an \emph{algebraically exact} dyadic phase thereafter.
 
\newcommand{\lcell}[1]{\begin{tabular}[c]{@{}l@{}}#1\end{tabular}}
\newcommand{\ccell}[1]{\begin{tabular}[c]{@{}c@{}}#1\end{tabular}}

\begin{table*}[t]
\centering
\footnotesize
\renewcommand{\arraystretch}{1.35}
\setlength{\tabcolsep}{6pt}
\begin{tabular}{@{}l c c c c l@{}}
\toprule
Method
& \ccell{Online\\qubits}
& \ccell{Online non-Clifford\\cost (per use)}
& \ccell{Online\\$T$-depth}
& \ccell{Exact\\dyadic?}
& \lcell{One-time preparation\\(amortized over uses)} \\
\midrule
\lcell{Clifford+$T$ synthesis\\\cite{ross2016optimal,kliuchnikov2013fast}}
& $0$
& $3\log_2(1/\varepsilon)\ T$
& $O(\log\tfrac{1}{\varepsilon})$
& no
& N/A \\
\midrule
\lcell{Repeat-until-success\\\cite{paetznick2014repeat,bocharov2015efficient}}
& $1$ (n.d.)
& \ccell{${\sim}1.15\log_2(1/\varepsilon)\ T$\\(exp.)}
& $O(\log\tfrac{1}{\varepsilon})$
& no
& N/A \\
\midrule
\lcell{Phase-gradient catalyst\\\cite{gidney2018halving}}
& $b{+}1$
& \ccell{${\sim}2b$ Tof.\\(${\sim}8b\ T$)}
& \ccell{$O(b)$\\{}[$O(\log b)$]}
& yes
& \lcell{${\sim}O(b\log\tfrac{b}{\varepsilon})\ T$;\\not Clifford-cultivable} \\
\midrule
\textbf{This work}
& $2^b{+}1$
& \ccell{$2^b$ Tof.\\(${\sim}4{\cdot}2^b\ T$)}
& $\boldsymbol{O(1)}$
& yes
& \lcell{surface-code cultivation,\\Clifford-verifiable} \\

\midrule\midrule
\multicolumn{6}{@{}l@{}}{\textit{Concrete numbers for the $\sqrt{T}=Z^{1/8}$ gate ($b=3$),}}\\
\multicolumn{6}{@{}l@{}}{\textit{taking $\varepsilon=10^{-10}$ for the synthesis methods:}}\\
\midrule
Clifford+$T$ synthesis
& $0$
& ${\approx}100\ T$
& ${\approx}100$
& no
& N/A \\
\midrule
Repeat-until-success
& $1$ (n.d.)
& \ccell{${\approx}38\ T$\\(exp.)}
& \ccell{${\approx}38$\\(exp.)}
& no
& N/A \\
\midrule
Phase-gradient catalyst
& $4$
& \ccell{${\approx}7$ Tof. /\\$28\ T$}
& \ccell{${\approx}7$\\{}[${\approx}4$]}
& yes
& \lcell{synthesize constituent\\$R_z(2\pi/2^k)$} \\
\midrule
\textbf{This work}
& $9$
& \ccell{$8$ Tof. / $32\ T$\\(or $8\,\ket{\mathrm{CCZ}}$)}
& $\boldsymbol{O(1)}$
& yes
& \lcell{${\sim}10^{-6}$ leakage in\\${\sim}7$ attempts at $p{=}10^{-3}$} \\
\bottomrule
\end{tabular}
\caption{Resource comparison for implementing a fine dyadic phase $Z^{2^{-b}}$, separating the online cost paid per invocation from the one-time preparation cost of any reusable resource state. ``n.d.'' marks the non-deterministic measurement-and-retry structure of RUS. For the phase-gradient catalyst a single-qubit rotation is a $b$-bit addition \emph{controlled} by the rotation target, which costs roughly twice the bare adder in both Toffoli count and depth; the bracketed $O(\log b)$ $T$-depth is achievable with a logarithmic-depth (carry-lookahead) adder~\cite{gidney2018halving} at an increased Toffoli count. Toffoli-to-$T$ conversions use the measurement-assisted logical-AND of Ref.~\cite{gidney2018halving} ($4\,T$ per Toffoli). Our online $T$-depth is constant in $b$ via the GHZ/CNOT-tree fanout of Sec.~\ref{sec:protocol-catalyst}.}
\label{tab:comparison}
\end{table*}
 
\subsection{The competing methods}
 
\emph{Clifford$+T$ synthesis.} The optimal ancilla-free number-theoretic synthesis of a single-qubit $z$-rotation to accuracy $\varepsilon$ uses $3\log_2(1/\varepsilon)+O(\log\log(1/\varepsilon))$ $T$ gates~\cite{ross2016optimal,kliuchnikov2013fast}, an exponential improvement over the Solovay--Kitaev scaling~\cite{dawson2006solovay} but still logarithmic in the inverse target error. The method needs no ancilla and no resource state, but the rotation is \emph{approximate}: each use injects a coherent over- or under-rotation of size $\varepsilon$, and the full $T$ sequence must be executed in series on the data qubit, giving an online $T$-depth of $O(\log(1/\varepsilon))$. In logical computation, $\epsilon$ is usually taken as the logical error rate of the fault-tolerant protocol.
 
\emph{Repeat-until-success (RUS).} RUS circuits use a single ancilla, a short Clifford$+T$ gadget, and a measurement that heralds success; on failure a Clifford correction is applied and the gadget is repeated~\cite{paetznick2014repeat,bocharov2015efficient}. This trades determinism for a roughly threefold reduction in expected $T$-count, to ${\approx}1.26\log_2(1/\varepsilon)$~\cite{paetznick2014repeat} and ${\approx}1.15\log_2(1/\varepsilon)$ in the optimized Clifford$+T$ construction of Ref.~\cite{bocharov2015efficient}. The rotation remains approximate, and the variable number of rounds makes both the $T$-count and the latency random variables. 
 
\emph{The phase-gradient catalyst.} The phase-gradient state on $g$ qubits, $\ket{\nabla_g}=2^{-g/2}\sum_{k=0}^{2^g-1}e^{2\pi\ii k/2^g}\ket{k}$, is an eigenstate of the modular adder: controlled-adding a $b$-bit constant into the register kicks back an \emph{exact} dyadic phase onto the control and returns the state to itself, so it is genuinely reusable~\cite{kitaev1995quantum,kitaev2002classical,gidney2018halving}. This is the closest relative of our scheme, and the honest comparison is the most instructive. Online, the phase-gradient state is more compact in qubits than ours, but its per-rotation non-Clifford cost is larger than a naive count suggests. Because the rotation acts on a single qubit, the constant addition into the gradient register must be \emph{controlled} by that target qubit, which roughly doubles the bare-adder cost to ${\sim}2b$ Toffoli gates, with a $T$-depth set by the (now controlled) carry chain, on ${\sim}b{+}1$ qubits, versus our $2^b$ Toffoli gates on $2^b{+}1$ qubits. For large $b$ this remains an exponential online advantage in its favor; at $b=3$, however, the controlled adder (${\approx}7$ Toffoli, ${\approx}28\,T$, $T$-depth ${\approx}7$) is already comparable in non-Clifford count to our $8$ Toffoli ($32\,T$) and is deeper, so its only clear online edge there is the smaller register ($4$ versus $9$ qubits). Nevertheless, two structural differences cut the other way, as we now detail.
 
\subsection{Where our protocol compares favorably (and unfavorably)}
 
With these competing methods in mind, we list the following three comparisons that favor our protocol.
 
\paragraph{Exactness and error-budget independence (vs.\ synthesis and RUS).} Both synthesis and RUS inject a coherent rotation error $\varepsilon$ on \emph{every} invocation. In a computation that applies a fixed fine angle $M$ times, these errors can add coherently to ${\sim}M\varepsilon$, forcing $\varepsilon\lesssim\varepsilon_{\mathrm{target}}/M$ and raising the per-use cost to $3\log_2(M/\varepsilon_{\mathrm{target}})$ $T$ gates. Our catalyst injects \emph{zero} synthesis error on every use: the per-use cost ($8$ Toffoli, or $32\ T$, for $\sqrt{T}$) is flat in both $M$ and the target precision. Concretely, at the $\varepsilon=10^{-10}$ budget appropriate once logical leakage reaches $10^{-7}$ or below, ancilla-free synthesis costs ${\approx}100\ T$ per use and RUS ${\approx}38$ expected $T$ per use, whereas our exact invocation costs $32\ T$. Our scheme therefore undercuts even the optimized RUS $T$-count \emph{and} removes its approximation error and its nondeterminism simultaneously.
 
\paragraph{Constant online $T$-depth (vs.\ all three).} This is the differentiator that no competitor matches. Synthesis and RUS have online $T$-depth $O(\log(1/\varepsilon))$; the ripple phase-gradient adder has $T$-depth $O(b)$, reducible only to $O(\log b)$ with a look-ahead adder~\cite{draper2006logarithmic}. Our invocation has $T$-depth $O(1)$ \emph{independent of $b$}, because the controlled-$U_n$ Toffoli gates are applied in parallel through the GHZ/CNOT-tree fanout of Sec.~\ref{sec:protocol-catalyst}. In a reaction-limited architecture, where the logical cycle and decoding latency dominate wall-clock runtime, a constant-depth fine rotation is a genuine structural advantage. The natural favorable workload is a latency-critical inner loop that applies the same fine phase repeatedly in series, for example a Trotterized phase evolution at a fixed dyadic step, or a fixed subroutine invoking $\sqrt{T}$ on its critical path. In these cases, our flat critical path beats the logarithmic growth of synthesis and the (at best) logarithmic-in-$b$ growth of the phase-gradient adder.
 
\paragraph{Fault-tolerant preparability (vs.\ the phase-gradient catalyst).} The decisive contrast with the phase-gradient state lies not online but in preparation. The phase-gradient state is an eigenstate of the \emph{non-Clifford} adder. Equivalently, it factorizes as $\bigotimes_{k=1}^{g} Z^{2^{-k}}\ket{+}$, so preparing it directly requires applying fine $z$-rotations, including the finest, $Z^{2^{1-g}}$, which is the very gate one is trying to produce. Its only exact verification is a measurement in the eigenbasis of addition, which is itself non-Clifford. Consequently there is no known low-distance-to-high-distance \emph{cultivation} of the phase-gradient state from postselected Clifford operations: its fault-tolerant preparation reduces either to synthesizing the constituent fine rotations to the target precision, that is, a one-time but $\varepsilon$-dependent Clifford+$T$ cost, ${\sim}O(b\log(b/\varepsilon))$ $T$~\cite{gidney2018halving}. Our catalyst, by contrast, is an eigenstate of a \emph{Clifford} circuit $U_n$ (Sec.~\ref{sec:protocol-catalyst}). It is therefore verifiable by a single \emph{exact} logical-$U_9$ phase-estimation check and cultivable to a logical leakage of $10^{-6}$--$10^{-7}$ on the rotated surface code using \emph{only} postselected Clifford operations and a constant magic budget (Sec.~\ref{sec:simulation}). When the fine catalyst must itself be produced fault-tolerantly to cultivation-grade fidelity from scratch, with no pre-existing high-precision fine-rotation oracle to bootstrap from, our route supplies a concrete, Clifford-only surface-code pathway, while the phase-gradient state inherits a preparation cost that scales with the very precision one is trying to reach.

The catalyst is not a universal replacement, and it is worth being plain about where it loses. Its main drawback is the size of the resource state: holding the catalyst for the phase $2\pi/2^{b+1}$ takes $2^b+1$ logical surface-code blocks (nine for $\sqrt{T}$), so the qubit overhead, and with it the online Toffoli count, grows exponentially with the fineness $b$. Per-gate synthesis and RUS, by contrast, use no resource state and at most a single ancilla, and their cost does not depend on $b$ at all; it is set by the target accuracy $\varepsilon$ and scales only as $\log(1/\varepsilon)$. The conventional phase-gradient state is also more economical at large $b$, storing the same phase in $O(b)$ qubits. In effect, our construction trades the $\varepsilon$-dependence of synthesis for a $b$-dependence: the rotation is exact and costs nothing for precision, but its footprint becomes prohibitive for very fine angles, so the protocol is practical only for modest $b$. It is, moreover, tied to one angle at a time: a workload spread over many distinct, arbitrary rotations is better served by synthesis, since cultivating a separate catalyst for each angle is impractical. The features that define its niche are correspondingly modest but concrete. The catalyst is reusable, so its cultivation is a one-time cost amortized over all later invocations rather than paid per gate; the rotation it applies is exact, so it injects none of the approximation error $\varepsilon$ that synthesis and RUS incur on every use, and that error neither builds up over a long computation nor forces recompilation as the error budget tightens; and its online non-Clifford depth is constant in $b$. The protocol is thus best suited to a small number of fixed, moderately fine dyadic phases that are reused many times. A detailed, algorithm-by-algorithm comparison of the various fine-phase protocols, weighing these advantages against the qubit overhead in realistic settings, is left to future work.

 \section{Conclusion and Outlook}
 
We have generalized magic-state cultivation to the setting of logical \emph{catalyst} states, cultivating an eigenstate of the high-period brickwork Clifford unitary $U_9$ whose eigenvalue $e^{\ii\pi/8}$ realizes an exact $\sqrt{T}$ phase. The entire protocol runs on blocks of surface code: a nine-qubit physical eigenstate is encoded into nine $\mathrm{Rot}(3)$ blocks and verified by a feedback-free semiclassical QPE measurement of the logical Clifford $\bar U_9$, assisted by GHZ fan-out, before the blocks are grown to $\mathrm{Rot}(7)$. Because the eigenphase verification is itself fault-tolerant, a single logical round suffices to reach the target leakage rate, a qualitative saving over the two rounds needed to cultivate single-block states such as $\ket{T}$. A hybrid tensor-network and stabilizer simulation confirms this behavior directly.
 
The most immediate extension is a fault-distance-5 protocol, adding a round of logical $U_9$ verification (again, plausibly a single round) and stabilizer measurement at the $\mathrm{Rot}(5)$ stage. We did not simulate it here for the limitation of classical resources, but the close numerical correspondence with $\ket{T}$ cultivation makes us confident it can drive the logical leakage to $\sim 10^{-9}$ or below with a $p=10^{-3}$ level physical error rate.
 
More broadly, our results show that cultivation can target the eigenstate of an intricate Clifford circuit, which raises the prospect of cultivating other \emph{structured} magic states that carry Clifford stabilizers. The $\ket{\mathrm{CCZ}}$ state, for instance, is stabilized by operators of the form $X\otimes CZ$ and could therefore in principle be cultivated directly; benchmarking an optimized $\ket{\mathrm{CCZ}}$ cultivation against the standard route of assembling it from four cultivated $\ket{T}$ states would be a natural and practically relevant next step.
 
The chief price of our construction, set against its constant online $T$-depth, is the $O(2^b)$ surface-code blocks needed to hold a catalyst for the phase $2\pi/2^{b+1}$, as quantified in Sec.~\ref{sec:application}. Reducing this overhead is the central open problem the present work poses. The natural target is the phase-gradient state, which stores the same phase resolution in only $b$ qubits but is an eigenstate of \emph{integer} modular addition on $\mathbb{Z}_{2^b}$, which is a non-Clifford operation. This places it beyond the Clifford-only verification used here. A promising route is to begin from a small code that supports a fault-tolerant logical adder, for example the quantum Reed--Muller code that admits transversal $T$ and $CNOT$ gates, and to verify the logical-adder eigenvalue with fanned-out ancillas, in direct analogy with our logical-$U_9$ check. Such a construction would open a cultivation pathway for phase-gradient states, and more generally for any catalyst state whose defining symmetry is cultivation-friendly.
 
Finally, embedding $\sqrt{T}$ and finer dyadic gates into useful quantum algorithms is an active area, and the catalytic, constant-depth, exactly-dyadic character of our gate is well matched to it. We expect that schemes for \emph{batched} cultivation, injection, and reuse of catalyst states, amortizing the preparation cost across many parallel invocations, will be an important ingredient in converting these gates into end-to-end algorithmic speedups.

\textit{Acknowledgments.} We thank Liyuan Chen for sharing unpublished result of a forthcoming work. YX acknowledges support by the NSF through the grant OAC-2118310, and the Gordon and Betty Moore Foundation's EPiQS Initiative, Grant GBMF10436.
XW is supported by the U.S. Department of Energy through Award Number DE-SC0023905.
The computation was performed on high-performance computing clusters supported by the Gordon and Betty Moore Foundation's EPiQS Initiative, Grant GBMF10436, and the NSF through the AI Research Institutes program (Award No. DMR-2433348).
 
\textit{Code availability.} The code used to simulate our protocol is available at ~\cite{xiaowang2026}.
 
\bibliography{main}
 
\appendix
\section{Unitary encoding circuit for $\rm rot(3)$ surface codes}\label{app:scencode}
 
The cultivation protocol begins by preparing the nine-qubit state
\(\ket{\psi_9}\) and encoding its \(j\)-th qubit into the \(j\)-th
distance-three $\rm rot(3)$ surface code block. Here we spell out the detailed unitary encoding circuit. We note that the same circuit is used in the hand-off simulation.
 
For concreteness, we label the nine qubits in the \(\mathrm{Rot}(3)\) patch in a row-major order
\begin{equation}
\begin{matrix}
1 & 2 & 3\\
4 & 5 & 6\\
7 & 8 & 9 ,
\end{matrix}
\label{eq:d3-layout}
\end{equation}
where $1$ is the logical register and the rest eight qubits, from $2$ to $9$, are the syndrome register, which are initialized in $\ket{0}$.
We then apply
\begin{equation}
    H_2 H_3 H_6 H_8,
    \label{eq:encoder-H}
\end{equation}
which switches the basis for the four $X$ syndrome registers. This is followed by the CNOT sequence
\begin{equation}
\begin{aligned}
    &(6\to5),\quad
    (2\to1),\quad
    (8\to7),\quad
    (2\to5),\quad
    (1\to4),\\
    &(6\to9),\quad
    (5\to8),\quad
    (4\to7),\quad
    (3\to2).
\end{aligned}
\label{eq:encoder-cnot}
\end{equation}
We denote the collection of the Hadamard and CNOT gates in Eqs.~\eqref{eq:encoder-H} and \eqref{eq:encoder-cnot} as \[U_{\rm enc}.\]
With the convention in Eq.~\eqref{eq:d3-layout}, the logical operators are
chosen as
\begin{equation}
    \bar Z = Z_1 Z_2 Z_3,
    \qquad
    \bar X = X_1 X_4 X_7 .
    \label{eq:d3-logicals}
\end{equation}
 
In the physical protocol, the preparation of \(\ket{\psi_9}\), the encoder
CNOTs, and the subsequent logical checks are part of the noisy front-end
circuit.  In the hybrid simulation of Sec.~\ref{sec:simulation}, this
front-end is handled by tensor-network evolution.  After postselection, the
state is exported as a nine-logical-qubit handoff together with the sampled
stabilizer-register sectors of the nine blocks. The subsequent stabilizer
simulation re-applies Eqs.~\eqref{eq:encoder-H} and \eqref{eq:encoder-cnot}
as a noiseless representation of that already-prepared handoff state, not as
an additional noisy physical re-encoding step.

\section{Details of the logical measurement circuit of $U_9$}\label{app:explicitcnots}
 
\begin{table*}[t]
    \begin{tabular}{cc}
    \toprule
    Set & Directed block pairs $(s,t)$ \\
    \midrule
    $E_1$ & $(1,2),(3,4),(5,6),(7,8),(2,3),(4,5),(6,7),(8,9)$ \\
    $E_2$ & $(1,3),(1,4),(2,4),(2,5),(3,5),(2,6),(3,6),(4,6),(4,7),(5,7),(4,8),(5,8),(6,8),(6,9),(7,9)$ \\
    $E_4$ & $(1,5),(2,6),(1,7),(3,7),(1,8),(2,8),(4,8),(2,9),(3,9),(5,9)$ \\
    $E_8$ & $(1,9)$ \\
    \bottomrule
 
    \end{tabular}
    \caption{Simplified CNOT implementations for $U_9^{2}$, $U_9^{4}$, $U_9^{8}$ given by the edges in $E_{2,4,8}$.}
    \label{tbl:cnotedges}
\end{table*}
\begin{figure*}[t]
\centering
\begin{quantikz}[column sep=0.27cm, row sep=0.16cm]
\lstick{$a_{1,1}\quad\ket{+}$}   & \ctrl{1} & \ctrl{2} & \ctrl{12} & \qw & \qw & \qw & \ctrl{1} & \ctrl{2} &\qw & \meter{X} \\
\lstick{$a_{1,2}\quad\ket{0}$}   & \targ{}  & \qw      & \ctrl{11} & \qw & \qw & \qw & \targ{}  & \qw      & \meter{}\\
\lstick{$a_{1,3}\quad\ket{0}$}     & \qw      & \targ{}  & \ctrl{10} & \qw & \qw & \qw & \qw      & \targ{}  & \meter{} \\
\lstick{$a_{2,1}\quad\ket{+}$}  & \ctrl{1} & \ctrl{2} & \qw & \ctrl{9} & \qw & \qw & \ctrl{1} & \ctrl{2} & \gate{S^\dagger} & \meter{X} \\
\lstick{$a_{2,2}\quad\ket{0}$}      & \targ{}  & \qw      & \qw & \ctrl{8} & \qw & \qw & \targ{}  & \qw      & \meter{}\\
\lstick{$a_{2,3}\quad\ket{0}$}     & \qw      & \targ{}  & \qw & \ctrl{7} & \qw & \qw & \qw      & \targ{}   & \meter{}\\
\lstick{$a_{3,1}\quad\ket{+}$}  & \ctrl{1} & \ctrl{2} & \qw & \qw & \ctrl{6} & \qw & \ctrl{1} & \ctrl{2} &  \gate{T^\dagger} & \meter{X} \\
\lstick{$a_{3,2}\quad\ket{0}$}     & \targ{}  & \qw      & \qw & \qw & \ctrl{5} & \qw & \targ{}  & \qw      & \meter{}\\
\lstick{$a_{3,3}\quad\ket{0}$} & \qw          & \targ{}  & \qw & \qw & \ctrl{4} & \qw & \qw      & \targ{}  & \meter{}\\
\lstick{$a_{4,1}\quad\ket{+}$} & \ctrl{1} & \ctrl{2} & \qw & \qw & \qw & \ctrl{3} & \ctrl{1} & \ctrl{2} &  \gate{\sqrt{T}^\dagger} & \meter{X} \\
\lstick{$a_{4,2}\quad\ket{0}$}       & \targ{}  & \qw      & \qw & \qw & \qw & \ctrl{2} & \targ{}  & \qw      & \meter{}\\
\lstick{$a_{4,3}\quad\ket{0}$}       & \qw      & \targ{}  & \qw & \qw & \qw & \ctrl{1} & \qw      & \targ{}   & \meter{}\\
\lstick{$\Dcal\quad\quad$}       & \qw      & \qw      & \gate{\mathsf{L}_{A_1}(E_8)} & \gate{\mathsf{L}_{A_2}(E_4)} & \gate{\mathsf{L}_{A_3}(E_2)} & \gate{\mathsf{L}_{A_4}(E_1)} & \qw & \qw & \qw &\qw &\qw
\end{quantikz}
\caption{One round of logical $U_9$ measurement circuit with deterministic feedback.}
\label{fig:logicalu9circ}
\end{figure*}
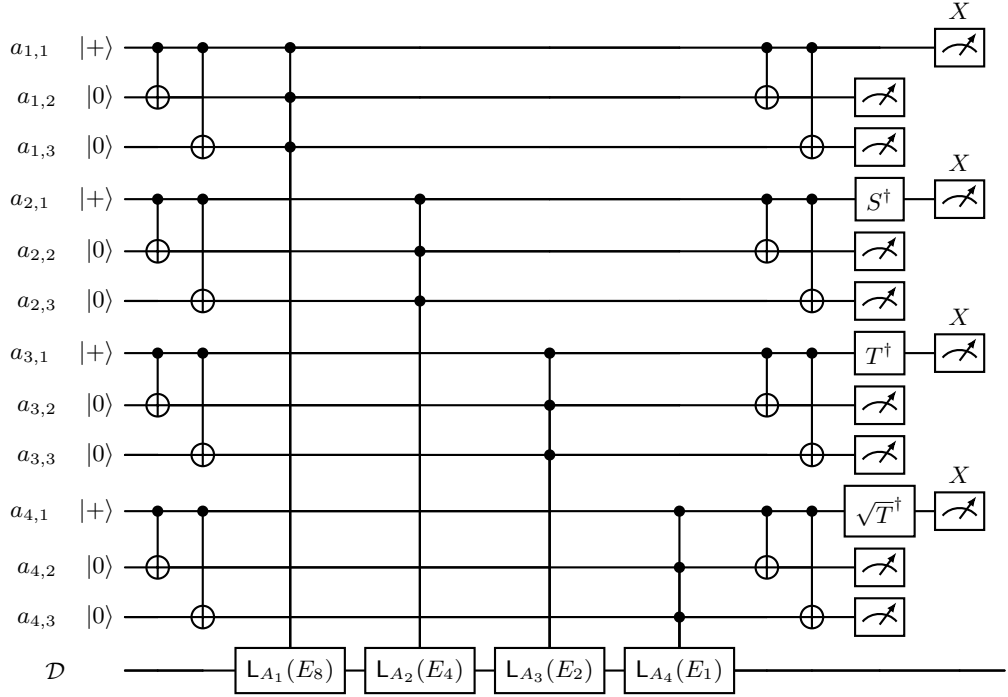
 
In this appendix, we spell out the details of one round of logical $U_9$ measurement. For concreteness, we fix the label of each qubit in this part of the protocol. We label the 9 data qubits in the $b$-th Rot(3) block by
\[
  D_b=(d_{b,1},d_{b,2},\ldots,d_{b,9}),\qquad b=1,\ldots,9.
\]
 The aggregate data register is
\[
  \Dcal=(D_1,D_2,\ldots,D_9), \qquad |\Dcal|=81.
\]
We designate the first qubit of each block as the logical register before encoding. That is, the physical eigenstate $\ket{\psi_9}$ is supported on $d_{b,1}$, $b=1,\ldots,9$.
In this way, $d_{b,2},\ldots,d_{b,9}$ are the sampled stabilizer registers before encoding. 
A logical-$U$ measurement round uses four 3-qubit GHZ state registers
\[
 A_1=(a_{1,1},a_{1,2},a_{1,3}),\ldots,A_4=(a_{4,1},a_{4,2},a_{4,3}).
\]
Each GHZ register is fanned out from a single $\ket{+}$ state on $a_{j,1}$, $j=1,\ldots,4$.
 
The logical-$U$ measurement is achieved using controlled-logical-$U_9$ gates with powers of $U_9$, see Eq.~\eqref{eq:U9-circuit}. Here every CNOT gate in $\overline{U_9}$ is implemented via transversal CNOT (tCNOT), which is a collection of physical CNOTs:
\begin{equation}
    \overline{\mathrm{CNOT}_{c\to d}}=\prod_{i=1}^9\mathrm{CNOT}_{d_{c,i}\to d_{d,i}}.
\end{equation}
To decrease circuit depth of logical measurement, rather than directly applying controlled-$U_9$ multiple times to reach the high powers, we find that higher powers of $U_9$ itself admit simplified implementations, which we summarize in Table~\ref{tbl:cnotedges}.

The circuit applies the sets of controlled-tCNOT gates in the order $E_8,E_4,E_2,E_1$, controlled by $A_1,A_2,A_3,A_4$ respectively.
For an edge $(s,t)\in E_p$ and an ancilla register $A_k=(a_{k,1},a_{k,2},a_{k,3})$, the implementation expands the block-level operation into nine physical Toffoli gates:
\begin{equation}\label{eq:lagates}
  C_{A_k}\mathsf{L}(E_p)
  \equiv \prod_{(s,t)\in E_p}\prod_{m=0}^{2}\prod_{\ell=1}^{3}
      \mathrm{CCX}\!\bigl(a_{k,\ell},\ d_{s,3m+\ell};\ d_{t,3m+\ell}\bigr),
\end{equation}
where $\mathsf{L}(E_p)$ represents the collection of physical CNOT gates in the transversal implementation of the CNOT gates in $E_p$ and $C_{A_k}\mathsf{L}(E_p)$ is the corresponding controlled gate via the GHZ register $A_k$. The product order matches the nested loops in the code: edge order first, then $m=0,1,2$, then lane $\ell=1,2,3$ which can be implemented in parallel.
Thus one logical-$U$ measurement round contains
\[
  9(|E_8|+|E_4|+|E_2|+|E_1|)=9(1+10+15+8)=306
\]
physical Toffoli gates.
 
After all the gates in Eq.~\eqref{eq:lagates} are implemented, every ancilla register is uncomputed and the clean-up qubits $a_{k,2}$ and $a_{k,3}$ for $k=1,\ldots,4$ are measured out.
 
The remaining 4 registers $a_{j,1}$, $j=1,\ldots,4$, undergoes a semi-classical QFT circuit with deterministic feedback, which applies physical $S^\dagger$, $T^\dagger$ and $\sqrt{T}^\dagger$ to the registers $a_{2,1}$, $a_{3,1}$ and $a_{4,1}$, respectively. The circuit diagram of the entire logical-$U_9$ measurement is presented in Fig.~\ref{fig:logicalu9circ}.

\begin{table}[t]
  \centering
  \begin{tabular}{c c c c}
    \hline
    $q$ & $X_q$ & $Y_q$ & $Z_q$ \\
    \hline
    $1$ & $2.30$ & $2.30$ & --   \\
    $2$ & $1.70$ & $1.73$ & $1.00$\\
    $3$ & $1.71$ & $1.71$ & $1.50$\\
    $4$ & $1.71$ & $1.71$ & $1.33$\\
    $5$ & $2.00$ & $1.33$ & $1.75$\\
    $6$ & $1.33$ & $1.71$ & $1.67$\\
    $7$ & $1.50$ & $1.71$ & $1.67$\\
    $8$ & $1.00$ & $1.33$ & $1.71$\\
    $9$ & --    & $1.88$ & $1.88$\\
    \hline
  \end{tabular}
  \caption{Expected syndrome weight (Hamming distance of the four-bit
  logical-$U_9$ read-out from the ideal string $1000$), conditioned on a
  nontrivial read-out, for each single-qubit Pauli fault on the physical
  catalyst $\ket{\psi_9}$. Dashes mark the two faults ($Z_1$, $X_9$) that act as
  a global phase and never produce a nontrivial read-out. The
  Born-probability-weighted average over all $27$ faults is $\bar w=1.654$.}
  \label{tbl:u9syndrome}
\end{table}

\section{Fault-tolerance of the \texorpdfstring{$U_9$}{U9} eigenspace}\label{app:u9ft}
 
This appendix details the fault analysis behind the effective logical
fault distance $\mathbf{f}_{\mathrm{logical}}\approx2.654$ quoted in
Sec.~\ref{sec:protocol-cultivation-verification}. We work with the bare physical catalyst
$\ket{\psi_9}$ of Eq.~\eqref{eq:psi9}, the $e^{\ii\pi/8}$ eigenstate of $U_9$ generated as the length-$16$ orbit of
$\ket{100000000}$ in Eq.~\eqref{eq:psi9}, and ask how a single-qubit Pauli fault $E$ is reported by
the actual read-out used in the protocol.
 
The read-out is the \emph{feedback-free} semiclassical phase estimation of
Sec.~\ref{sec:protocol-cultivation-verification} and Fig.~\ref{fig:logicalu9circ}: four ancillas control
$U_9^{8},U_9^{4},U_9^{2},U_9^{1}$, and before the $X$-basis measurement each
receives a \emph{fixed} phase correction, namely $I$, $S^\dagger$, $T^\dagger$, and
$\sqrt{T}^\dagger$, respectively, which are chosen to map the target eigenphase to the
string $1000$. Because these corrections are deterministic rather than
conditioned on earlier outcomes, the four bits are measured in a fixed product
basis. Writing the read-out string and the power index as four-bit integers
$s=s_1s_2s_3s_4$ and $t=t_1t_2t_3t_4$ (with $s_1$ the most significant bit,
controlling $U_9^{8}$), the outcome probability on input $E\ket{\psi_9}$ is
\begin{align}
  &P(s)=\bigl\lVert M_s\,E\ket{\psi_9}\bigr\rVert^{2},\nonumber\\  
  &M_s=\frac{1}{16}\sum_{t=0}^{15}(-1)^{s\cdot t}\,
       e^{-\ii\left(\frac{\pi}{2}t_2+\frac{\pi}{4}t_3+\frac{\pi}{8}t_4\right)}U_9^{\,t},
  \label{eq:eigenphase-projector}
\end{align}
with $s\cdot t=\sum_k s_kt_k$. Crucially, the fixed corrections are tuned so
that the \emph{accepted} outcome reproduces the eigenphase projector exactly,
\begin{equation}
  M_{1000}=\Pi_1\equiv\frac{1}{16}\sum_{t=0}^{15}e^{-2\pi\ii t/16}\,U_9^{\,t},
  \label{eq:eigenphase-accept}
\end{equation}
the projector onto the $e^{\ii\pi/8}$ eigenspace ($1000$ being the bit-reversal
of $0.0001_2$) Since $U_9$ has period $16$ its eigenvalues lie exactly on the
four-bit grid, so this read-out has no spectral leakage. We define the
\emph{syndrome weight} of a shot as the Hamming distance between its read-out
and $1000$, so that an undetected fault has weight $0$ and a maximally flagged
one has weight $4$.
 
Two structural facts, both consequences of Eq.~\eqref{eq:eigenphase-accept} and
hence \emph{independent} of how the rejected outcomes are read out, make a
single round powerful. First, accepting $1000$ projects $E\ket{\psi_9}$
\emph{exactly} into the $e^{\ii\pi/8}$ eigenspace; under the leakage metric of
Eq.~\eqref{eq:shot-fidelity} the post-measurement state is then a legitimate
catalyst, and any component that has leaked to a different eigenphase is
\emph{certain} to register a read-out $\neq1000$. Hence no single-qubit fault
produces undetected leakage: masking a leak requires, in addition to the
physical fault, measurement faults that forge $1000$. (Components that return
$1000$ while differing from $\ket{\psi_9}$ merely populate other vectors of the
degenerate $e^{\ii\pi/8}$ eigenspace, which is innocuous for catalysis as noted
in Sec.~\ref{sec:protocol-cultivation-verification}.) Second, the only
single-qubit faults that \emph{always} return $1000$ are $Z_1$ and $X_9$: qubit
$1$ is set in every word of the orbit, so $Z_1\ket{\psi_9}=-\ket{\psi_9}$; and
$X_9$ coincides with $U_9^{8}=\mathrm{CNOT}_{1\to9}$ on the orbit, so
$X_9\ket{\psi_9}=U_9^{8}\ket{\psi_9}=-\ket{\psi_9}$. Both act as a global phase
and leave the catalyst invariant.
 
Evaluating Eq.~\eqref{eq:eigenphase-projector} for all $27$ faults
$\{X_q,Y_q,Z_q\}_{q=1}^{9}$, we find that two ($Z_1$, $X_9$) are undetectable
global phases, ten are flagged with certainty, and the remaining fifteen are
flagged with probability between $0.44$ and $0.94$.
Table~\ref{tbl:u9syndrome} lists, for each fault, the expected syndrome weight
conditioned on a nontrivial read-out. Averaging over the $27$ faults,
conditioned on a nontrivial read-out and weighted by the Born probabilities,
gives the mean syndrome weight
\begin{equation}
  \bar w \;=\; 1.654 ,
  \label{eq:wbar}
\end{equation}
so that masking a leaked single-qubit fault costs on average $1.654$
measurement faults and $\mathbf{f}_{\mathrm{logical}}=1+\bar w\approx2.654$
in Eq.~\eqref{eq:f-logical}. The detection probabilities and the two blind spots
follow solely from Eq.~\eqref{eq:eigenphase-accept} and are therefore exact. The
diagonal $Z$ faults are special: being diagonal, they preserve the orbit, so
each $Z_q$ is either an undetectable global phase ($q=1$) or, when it leaks,
fully detected. The $X$ and $Y$ faults can scatter $\ket{\psi_9}$ into other
$U_9$ orbits sharing the eigenphase and account for the partial read-outs.

\begin{figure}[t]
    \centering
    \includegraphics[width=\linewidth]{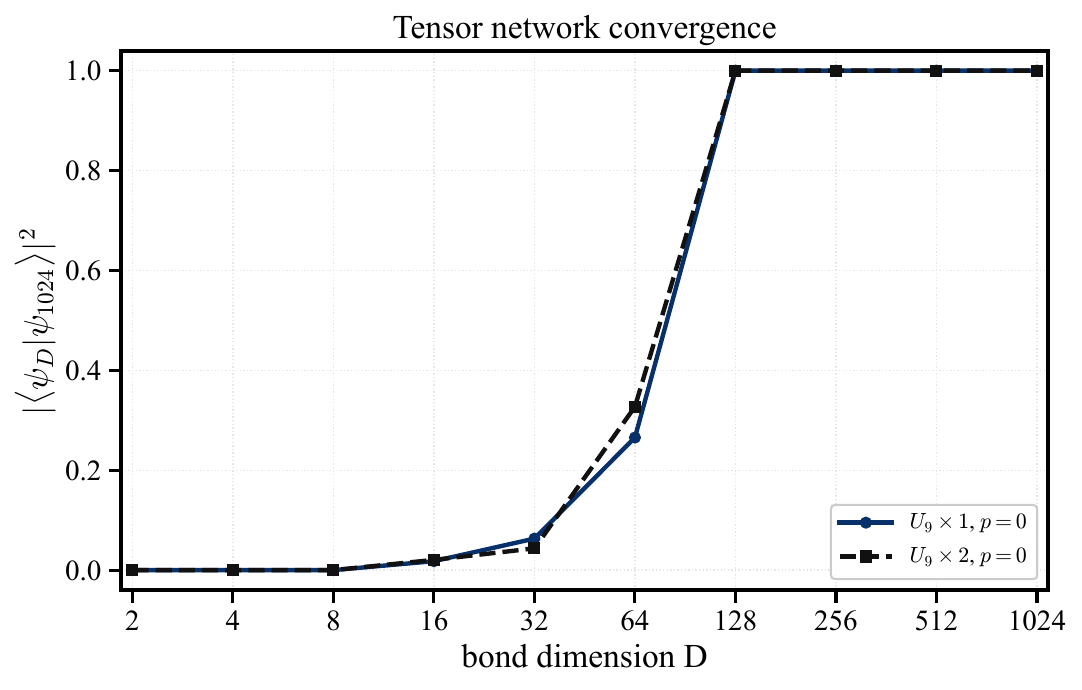}
    \caption{State vector convergence of the tensor network simulation as the maximum bond dimension is varied from $2^1$ to $2^{10}$. 
    The reference state is obtained with maximum bond dimension $2^{10}=1024$. 
    The horizontal axis is shown on a logarithmic scale in the bond dimension.}
    \label{figtnconvergence}
\end{figure}
 
\begin{figure*}[t]
    \centering
    \includegraphics[width=\linewidth]{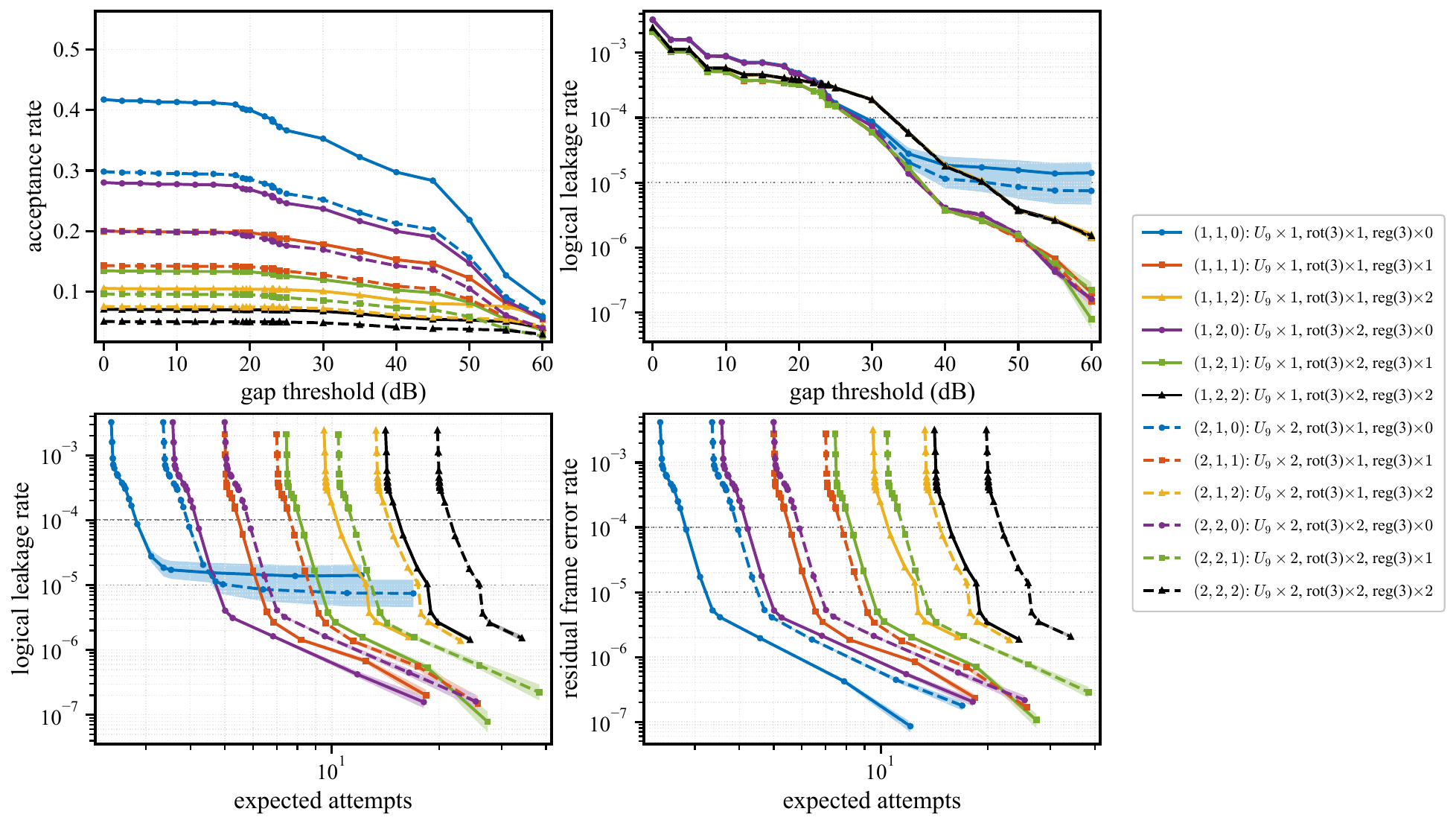}
    \caption{Detailed result of all different growth protocols with $N_1=1,2$, $N_2=1,2$ and $N_3=0,1,2$. (Top left) Acceptance rate versus gap threshold. (Top right) Logical leakage rate versus gap threshold. (Bottom left): Logical leakage rate versus expected attempts. (Bottom right) Backend residual frame error rate (cf. Eq.~\eqref{eq:residual-frame}) versus expected attempts.}
    \label{fig:fullresult}
\end{figure*}

\section{Convergence of the tensor-network simulation}\label{app:tnconverge}
 
Here we describe how the tensor network algorithm is used to generate the front end state vector.
The program is based on the \texttt{iTensor} library~\cite{itensor}.
As discussed in Sec.~\ref{sec:simulation-strategy}, 
the front end circuit contains hundreds of non-Clifford physical gates and therefore a dense simulation over the full Hilbert space is infeasible on a classical computer.
Using the tensor network algorithm, 
we construct the front end wave function as a matrix product state with finite bond dimension at a prescribed wave function accuracy, 
which produces the backend input for the Stim simulation.
For one round of logical $U_9$ measurement, 
the full front end circuit contains the $81$-qubit data register $\Dcal$ and $12$ ancilla qubits, 
giving $93$ qubits in total.
For two rounds of logical $U_9$ measurement, 
a literal implementation with a fresh ancilla register in each round would contain the $81$-qubit data register and $24$ ancilla qubits, 
giving $105$ qubits in total.
However, in the real tensor network simulation, 
we instead reset and reuse the measured ancilla qubits between the two rounds, 
so the number of active qubits remains $93$, 
which reduces the resource cost of the tensor network simulation.
 
We first set the stochastic gate noise rate to $p=0$ for a convergence test.
With the SVD cutoff for the discarded singular value two norm set to $10^{-10}$,
we run state vector simulations with several different upper bounds on the bond dimension.
Taking the state vector with maximum bond dimension $1024$ as the reference state,
we verify that the full state vectors used to obtain the backend input converge once the maximum bond dimension is larger than $128$ in both the settings with one and two rounds of logical $U_9$ verification,
as shown in Fig.~\ref{figtnconvergence}.
Furthermore,
the depolarizing channel described in Eq.~\eqref{eq:depol-channel} acts locally after local gates.
At the physical error rate used in the main simulations,
$p=10^{-3}$,
only a small number of stochastic Pauli faults is expected to occur across the hundreds of local gate locations in the full front end circuit.
Together,
these two facts suggest that the sampled stochastic faults should not cause a substantial statistical increase in the entanglement entropy or in the bond dimension required to represent the front end state.
We therefore use the state vector convergence at $p=0$ to estimate the bond dimension needed for the noisy sampling.
In the final simulations,
we choose $256$ as the maximum bond dimension.
We then apply stochastic Pauli operators according to the noise model in Eq.~\eqref{eq:depol-channel} in the tensor network simulation.
By repeatedly generating $2000$ front end samples,
we construct the state vector pool used for the Stim backend simulation.

\section{Comparison of growth strategies}\label{app:growth}
In this appendix, we present the remaining growth protocols with different $N_2$ and $N_3$ measurement rounds for $\rm Rot(3)$ and $\rm Reg(3)$ measurements. Just as the cases presented in the main text, each protocol is simulated with 2000 front-end shots using tensor-network based state-vector simulation, and $2\times 10^6$ backend shots using Stim and complementary-gap decoding, where we set the maximum gap threshold to 60dB.
 
We plot the detailed simulation result in Fig.~\ref{fig:fullresult} with different gap thresholds all the way to 60dB, and convert the acceptance rate to expected attempts. First, we see that $(N_2,N_3)=(1,0)$ fails to reach the logical leakage rate around $10^{-6}$ since the stabilizer fault distance does not reach 3. Second, $N_2+N_3=2$ is enough to achieve $10^{-7}$ level logical leakage rate. Adding more rounds of stabilizer measurements simply adds to more expected attempts with no meaningful improvement over the ones with $N_2+N_3=2$. In addition, stabilizer measurement at $\rm Rot(3)$ is in general more efficient than $\rm Reg(3)$. Finally, as noted in the main text, adding another round of logical $U_9$ measurement does not add to the quality of the final logical catalyst state, where the errors introduced by the logical QPE circuit likely break even with the added benefit of another logical check.

\end{document}